\documentclass[twocolumn,pra,aps]{revtex4}
\usepackage{psfrag}
\usepackage{epsfig}
\usepackage{amsmath}
\usepackage{amssymb}
\usepackage{bbm}
\usepackage{pifont}
\usepackage{psfrag}
\usepackage{ulem}
\usepackage{setspace}
\newcommand{\be}{\begin{equation}}
\newcommand{\ee}{\end{equation}}

\newcommand{\tr}{\textrm{tr}}
\newcommand{\im}{\text{i}}
\newcommand{\adop}{\hat{a}^{\dagger}}
\newcommand{\aop}{\hat{a}}

\newcommand{\Eop}{\hat{E}}

\newcommand{\bdop}{\hat{b}^{\dagger}}
\newcommand{\bop}{\hat{b}}

\newcommand{\rhos}{\hat{\rho}_{\rm S}}

\newcommand{\hc}{\text{H.c.}}

\newcommand{\ie}{{\it i.e.}}
\newcommand{\eg}{{\it e.g.}}

\newcommand{\bx}{\bold{x}}

\newcommand{\bk}{\bold{k}}
\newcommand{\br}{\hat{r}}

\begin{document}

\title{Master equation approach to optomechanics with arbitrary dielectrics}

\author{Anika C. Pflanzer}
\email{anika.pflanzer@mpq.mpg.de}
\author{Oriol Romero-Isart }
\author{J. Ignacio Cirac}
\affiliation{Max-Planck-Institut f\"ur Quantenoptik,
Hans-Kopfermann-Strasse 1,
D-85748, Garching, Germany.}

\begin{abstract}
We present a master equation describing the interaction of light with dielectric objects of arbitrary sizes and shapes. The quantum motion of the object, the quantum nature of light, as well as scattering processes to all orders in perturbation theory are taken into account. This formalism
extends the standard master equation approach to the case where interactions among different modes of the environment are considered. It yields a genuine quantum description, including a renormalization of the couplings and decoherence terms. We apply this approach to analyze cavity cooling of the center-of-mass mode of large spheres. Furthermore, we derive an expression for the steady-state phonon numbers without relying on resolved-sideband or bad-cavity approximations.
\end{abstract}

\maketitle

\section{Introduction}

In quantum optomechanics, light is used to cool and control the mechanical motion of massive objects in the quantum regime~\cite{Kippenberg08, Marquardt09, Karrai09, genes09}. These systems have potential applications in quantum information~\cite{ Stannigel10, Stannigel2011, Rabl10, Hammerer09}, metrology~\cite{cleland98, yang06, chiu08, Geraci10}, and can even be used in experiments testing the foundations of quantum mechanics~\cite{Bouwmeester03, RomeroIsart11b, RomeroIsart11c}. In the broad research area of cavity quantum optomechanics two classes of systems can be distinguished: the reflective case, realized in deformable Fabry-P\'erot resonators~\cite{kleckner06, gigan06} or microtoroidal cavities~\cite{Schliesser06}, and the dispersive case, like in the membrane-in-the-middle configuration~\cite{harris08, Jayich08, ni12} or in optically levitating nano-dielectrics~\cite{romeroisart10, romeroisart2011a, chang10, Barker10a, gieseler12}. In the latter, the dimension of the object along the cavity axis (\ie, the width of the membrane or the diameter of the nanosphere) is typically much smaller than the optical wavelength. This implies that the dielectric can be treated as a dipole with some induced polarizability~\cite{chang10,romeroisart2011a}. The problem is akin to that of single point particles, like atoms or ions, in the weak excitation regime. Thus, the theory and methods that have been developed in the context of laser cooling, trapping, and manipulation of single atoms and ions can be directly applied to optomechanical systems (see, \eg,~\cite{cirac92, leibfried03, raimond2001} for some expository articles). In particular, sideband-cooling techniques~\cite{Marquardt07, WilsonRae07} have been successfully employed to achieve the ground state in a nano-optomechanical system~\cite{chan11, verhagen12} (see also~\cite{Connell10, Teufel11}).

The control that is being achieved in dispersive quantum optomechanics opens up the challenge to explore the physics of larger objects. While this is certainly within experimental reach~\cite{tongcang11}, the existing quantum theories are not applicable since the dielectric object can no longer be considered as a simple dipole. In contrast, for sizes comparable or larger than the optical wavelength, multi-scattering processes within the dielectric have to be taken into account. As it is well-known from classical nano-photonics, they give rise to a modification of the forces experienced by the system, as well as other interesting phenomena~\cite{abajo07}.  

In this article, we present a quantum theory describing the interaction of light with the center-of-mass of non-absorbing dielectrics of arbitrary shapes and sizes. In particular, we derive a master equation for the motion of the particle and the cavity mode. This method considers the full scattering process by linking the coefficients of the master equation to the scattering matrix. This allows one to use the tools and techniques developed in the context of classical nano-photonics to determine the evolution of the quantum system. These include advanced numerical techniques, like the discrete-dipole approximation~\cite{purcell73}, the T-matrix method~\cite{Stroem75}, or, for some special geometrical shapes, even analytical solutions, like the 
so-called Mie solution~\cite{strattonbook, hulstbook, bohrenbook}. 

We apply this approach to a dielectric object in a high-finesse optical cavity and obtain an equation that is quadratic in the field operators of the center-of-mass and the cavity mode. For this general master equation, the steady-state phonon numbers are derived without relying on the common resolved-sideband or bad-cavity approximations. We illustrate this solution by analyzing the problem of laser-cooling in cavity optomechanics with optically levitating dielectric spheres of diameters comparable or even larger than the cavity mode wavelength. While ground-state cooling can be achieved for spheres much smaller than the wavelength, the minimal phonon numbers attainable for larger spheres oscillate around values of $n_{\rm min}\approx 500$. Note that this approach assumes that all photons are scattered into the bath modes, which is justified for objects that are not adapted to the cavity geometry, such as spheres or cylinders, but not for membranes~\cite{chang11}. 

The manuscript is organized as follows: In Sec.~\ref{sec:ass_ham}, we describe the system, listing the assumptions and defining the Hamiltonian. Following this, we present the main result of this manuscript in Sec.~\ref{sec:master}: a master equation describing the interaction between light and the motion of arbitrary dielectric objects. First, the effect of the presence of a dielectric on a free electromagnetic field is discussed in Sec.~\ref{sec:freefield}, where the corresponding scattering equations are solved. Based on this, we derive a general master equation describing the joint dynamics of the cavity mode and the center-of-mass motion of a dielectric in Sec.~\ref{sec:cavcm}. The description of the optomechanical setup is obtained by assuming the Lamb-Dicke regime and a strong driving field in Sec.~\ref{sec:optomech}.
In Sec.~\ref{sec:cooling} we investigate the possibility to cool dielectric objects deriving a general theory beyond the common approximations to determine phonon numbers in the steady state. Finally the optomechanical parameters for levitating spheres are determined in Sec.~\ref{sec:spheres}. In Sec.~\ref{sec:conclusion} we draw the conclusions and give an outlook on further directions.
In App.~\ref{classical} we review the classical solution of the scattering equations. Following this, in App.~\ref{sec:wigner}, we describe how the full spectrum of the electromagnetic field after interaction with a dielectric can be determined within an extended Wigner-Weisskopf approach. In App.~\ref{master_opto} we supplement the derivation of the master equation in the optomechanical case described in Sec.~\ref{sec:optomech}. The manuscript is rounded off in App.~\ref{limits} by a discussion of the small-particle limit.

\section{Physical model and Hamiltonian}\label{sec:ass_ham}

In this section we describe the system consisting of a dielectric object interacting with one or several confined electromagnetic modes. We discuss the assumptions that are taken and derive the complete Hamiltonian.
\subsection{Assumptions}
In this description of the interaction between a dielectric with a center-of-mass position $\br$ and a photonic field, the following assumptions are taken:
\begin{enumerate}
\item The object has a volume $V$, a density distribution $\rho$, and a mass $M=\rho V$. Note that the density distribution is assumed to be homogeneous for simplicity. In contrast to the common assumption, see \eg~\cite{romeroisart2011a}, we do not restrict the size of the dielectric to the sub-wavelength scale of the light field, but allow for arbitrary sizes.
\item The dielectric constant  $\epsilon_r$ is assumed to be homogeneous. The permeability of the object $\mu$ is chosen to be equal to the vacuum permeability, $\mu=\mu_0$, which is a good approximation for the dielectric objects we are mainly interested in. 
\item As shown in \cite{romeroisart2011a}, the center-of-mass (cm) mode of dielectrics at the micron-scale is decoupled from the vibrational ones. Hence, we will only consider the motion of the cm degree of freedom $\br$ and neglect its coupling to vibrational modes.
\item We assume the dielectric constant of the object to be real, \ie, no absorption effects are taken into account. In the language of scattering theory, this signifies that only elastic scattering processes are accounted for.
\item Throughout the first part of the paper, Sec.~\ref{sec:ass_ham}- Sec.~\ref{sec:master}, we assume the electromagnetic field to be scalar and neglect polarizations for a better readability of the equations. The derivations for polarizations can be carried out in full analogy. We use the results including polarizations in the analysis of cavity optomechanics with levitating spheres in Sec.~\ref{sec:spheres}.
\item We assume that all photons are scattered into the bath modes. This is a valid assumption for geometries that do not fit the cavity's geometry like, \eg, spheres, whereas for membranes the scattering into the cavity mode has to be taken into account~\cite{chang11}.
\end{enumerate}
\subsection{Hamiltonian}
The Hamiltonian consists of three parts, 
\be
\hat{H}_{\rm tot}=\hat{H}_{\rm M}+\hat{H}_{\rm L}+\hat{H}_{\rm LM}:\label{eq:Htot}
\ee
the cm motion of the dielectric is described by $\hat{H}_{\rm M}$, the energy of the electromagnetic field by $\hat{H}_{\rm L}$, and the interaction between the light and matter is given by $\hat{H}_{\rm LM}$. For the master equation description that we want to pursue in the proceeding, it is useful to divide the total Hamiltonian into
\be
\hat{H}_{\rm tot}=\hat{H}_{\rm S}+\hat{H}_{\rm B}+\hat{H}_{\rm BS},
\ee
where $\hat{H}_{\rm S}$ denotes the Hamiltonian describing the system, $\hat{H}_{\rm B}$ denotes the part describing the bath and $\hat{H}_{\rm BS}$ the coupling between the two. Each of these terms will be defined in the following.
\subsubsection{The kinetic energy}
The motion of the free untrapped dielectric is described by $\hat{H}_{\rm M}=\hat{p}^2/(2M)$, where $\hat{p}$ denotes the momentum operator of the cm coordinates in the direction we are interested in. While the dielectric object we investigate may have an arbitrary three-dimensional shape, we consider only its motion in one dimension. Due to the harmonicity of the trap, the coupling between the different directions can be neglected. Nevertheless, in many cases it might still be necessary to control the motion in the other directions, \eg, via feedback cooling~\cite{yin11}. In particular for linear or quadratic potentials, also the coupling to internal vibrational modes of the sphere can be neglected. In the absence of an additional external potential, the Hamiltonian can be diagonalized in the basis of the vibrational eigenmodes. Adding an external potential leads to some coupling between the cm degree of freedom and the vibrational modes. The frequency of the vibrational eigenmodes is roughly given by $\omega_n\propto n~c_{\rm sound}/R$, where $c_{\rm sound}$ denotes the sound velocity and $R$ the extension of the dielectric. For micron-scale objects it is several orders of magnitude larger than the trapping frequencies typically achieved for the cm degree of freedom. This enables one to adiabatically eliminate the vibrational modes merely leading to a negligible renormalization of the system's energy. A detailed discussion using a theory of quantum elasticity can be found in~\cite{romeroisart2011a}.
\subsubsection{The energy of the free electromagnetic field}
The energy of the free electromagnetic field is described by
\be
\hat{H}_{\rm L}=\frac{1}{2}\int d\bx\left[\epsilon_0\hat{E}_{\rm tot}^2(\bx)+\frac{\hat{B}_{\rm tot}^2(\bx)}{\mu_0}\right],
\ee
where $\epsilon_0$ denotes the vacuum permittivity, $\hat{E}_{\rm tot}$ the electric field and $\hat{B}_{\rm tot}$ the magnetic one. The total electromagnetic field can be divided into a part containing the continuous modes and one or several confined modes. The continuous part is defined as
\be
 \hat{E}_{\rm B}(\bx)=\frac{\im}{(2\pi)^{3/2}}\int d \bk\sqrt{\frac{\omega_{\bold{k}}}{2\epsilon_0}}(\aop_{\bold{k}}e^{-\im \bold{k} \bold{ x}}-\hc)\label{eq:freefield},
 \ee
where the label $B$ signifies that this continuum of plane-wave modes will generally be treated as a bath. The different modes are characterized by the annihilation (creation) operators $\aop_{\bk}$ ($\adop_{\bk}$) with a mode frequency $\omega_{\bk}$ and a wave vector $\bk$, where we will denote $k=|\bk|$. Note that we set $\hbar=1$ throughout the manuscript.
In the next step we define a confined mode of the electromagnetic field with annihilation (creation) operator $\aop_0$ ($\adop_0$), mode frequency $\omega_0$, mode volume $V_0$ and a mode profile given by $f(\bx)$. Typically it describes a mode in a cavity subject to some boundary conditions. We label this inhomogeneous part of the electromagnetic field $S$ (for system), it is given by
\be
\hat{E}_{\rm S}(\bx)=\im\sqrt{\frac{\omega_{0}}{2\epsilon_0V_{0}}}\left(\aop_{0}f(\bx)-\hc\right).\label{eq:einh}
 \ee
The extension to  several inhomogeneous modes can be achieved in an analogous fashion.

\subsubsection{Light-matter interaction}
The most interesting part of the Hamiltonian describes the interaction between the dielectric and the electromagnetic field. The response of the object's polarization is assumed to be linear to the electric field, which is fulfilled for the typical light intensities considered in this manuscript. The interaction Hamiltonian is given by
\be \label{eq:idielI}
\hat{H}_{\rm LM} =-\frac{1}{2}  \int_{V(\br)} d\bold{x} \hat{P}_{\rm tot}(\bold{x}) \hat{E}_{\rm tot}(\bold{x}),
\ee 
where $\hat{P}_{\rm tot}(\bold{x})$ is the object's polarization and the integration is performed over the volume of the dielectric $V$ with center-of-mass coordinate $\br$. Assuming $\hat{P}_{\rm tot}(\bx)= \alpha_p \hat{E}_{\rm tot}(\bx)$ and comparing the resulting relation between the polarization and the electric field for the macroscopic~\cite{jacksonbook} and microscopic case (see~\cite{romeroisart2011a} for a concise derivation), one obtains $\alpha_p=\epsilon_0\epsilon_r$ and
\be 
\hat{H}_{\rm LM} =-\frac{\epsilon_c\epsilon_0}{2} \int_{V(\hat{r})} d\bold{x}  \Eop_{\rm tot}(\bold{x})^2\label{eq:hamdi},
\ee 
where $\epsilon_c=3(\epsilon_r-1)/(\epsilon_r+2)$ is defined in terms of the relative dielectric constant $\epsilon_r$. Here, the cm is treated as an operator, such that Eq.~\eqref{eq:hamdi} gives the coupling terms between the object's position and the light field. 

Before describing the different contributions in detail, we reconsider the inhomogeneous mode $\Eop_{\rm S}$ that has been separated from the continuum, see Eq.~\eqref{eq:einh}. It describes one (or several) mode(s) that differs from the continuum. While in the specific setup of optomechanics with levitating spheres both the tweezer and the cavity field contribute, we describe this mode in general as the system mode.  Due to the high photonic occupation numbers that might occur in the presence of a strong driving field, it can be divided into a classical part and a quantum part by displacing the operators $\aop_0= \langle \aop_0\rangle+\aop'_0$ (note that we will omit the prime hereafter). This yields an additional contribution to the electromagnetic field given by 
\be
\begin{split}
\mathcal{E}_{\rm S}(\bx)&=\im\sqrt{\frac{\omega_{0}}{2\epsilon_0V_{0}}}\left(\alpha f(\bx)-\hc\right),\label{eq:einhcl}
\end{split}
\ee
where $\mathcal{E}_{\rm S}(\bx,t)$  is not an operator and describes the classical part of the light field with $\alpha=\langle \aop_0\rangle$, the square root of the photon number. 
Plugging $\Eop_{\rm tot}(\bx)=\Eop_{\rm S}(\bx)+\mathcal{E}_{\rm S}(\bx)+\Eop_{\rm B}(\bx)$
into Eq.~\eqref{eq:hamdi} leads to different contributions in the Hamiltonian $H_{\rm tot}$ of Eq.~\eqref{eq:Htot}. 
The Hamiltonian describing the system consisting of the inhomogeneous mode and the mechanical degree of freedom can be written as
\be
\begin{split}
\hat{H}_{\rm S}=\frac{\hat{p}^2}{2 M}+\omega_0\adop_0 \aop_0-\frac{\epsilon_c\epsilon_0}{2} \int_{V(\br)}d\bx\left(\mathcal{E}_{\rm S}(\bx)+\Eop_{\rm S}(\bx)\right)^2.\label{eq:Hsys}
\end{split}
\ee
The energy of the bath modes is given by 
\be
\hat{H}_{\rm B}=\int d\bk \omega_{\bk}\adop_{\bk}\aop_{\bk}+\hat{W}(\br),\label{eq:bath}
\ee
where $\hat{W}(\br)$ describes the interaction between different bath modes induced by the presence of the dielectric,  
\be
\hat{W}(\br)=-\frac{\epsilon_c\epsilon_0}{2} \int_{V(\br)}d\bx\Eop_{\rm B}^2(\bx).\label{eq:bathint}
\ee
The interaction between the system and the bath modes is denoted by
\be
\hat{H}_{\rm BS}=-\epsilon_c\epsilon_0 \int_{V(\br)}d\bx\left(\mathcal{E}_{\rm S}(\bx)+\Eop_{\rm S}(\bx)\right)\Eop_{\rm B}(\bx).\label{eq:hint}
\ee
The noninteracting part of the Hamiltonian describing the energy of the system and the bath is given by
\be
\hat{H}_0=\hat{H}_{\rm S}+\hat{H}_{\rm B}.
\ee
In quantum optics, the interaction between different bath modes, Eq.~\eqref{eq:bathint}, is commonly neglected.
In contrast, when describing the scattering of light from larger objects, interactions among the bath modes have to be taken into account, such that
it is no longer justified to neglect $W(\br)$.
$\hat{H}_0$ thus effects a coupling between different modes of the bath, such that the bath operators are not a diagonal basis anymore. We demonstrate in the following how this problem can be addressed and connected to a description within scattering theory.
\section{Master equation for arbitrary dielectrics}\label{sec:master}
We give a concise description of the two modes of the system we are interested in, the mechanical mode describing the center-of-mass motion of the dielectric and the cavity mode of the light. Therefore, we trace out the other modes of the electromagnetic field, the free modes. The typical quantum-optical approach to these systems is the method of Born-Markov master equations, where the bath is eliminated to derive a description exclusively for the system's dynamics. The Hamiltonian is split into a part describing the energy of the system and the bath, $\hat{H}_0$, and the interaction between the two, $\hat{H}_{\rm BS}$. For typical quantum-optical systems, $\hat{H}_0$ is diagonal in the bath operators $\aop_{\bk}$ as interactions among them are negligible, such that the transformation to the interaction picture is straightforward.
The difficulty we confront when describing the interaction between light and a dielectric sphere in a cavity larger than the wavelength is that due to the large number of scattered photons, interactions within the bath, given by Eq.~\eqref{eq:bathint}, have to be taken into account. This effects a Hamiltonian which is non-diagonal in the bath operators $\aop_{\bk}$. 

The strategy to approach this problem is to first solve the equations of motion, effected by the interaction with the dielectric, for the bath operators. Connecting these expressions to the Lippmann-Schwinger equation for the scattering process of a single photon, we give the solution for the bath operators containing the full scattering interaction in Sec.~\ref{sec:freefield}. Subsequently, we derive the master equation in Sec.~\ref{sec:cavcm} describing the cavity mode and the center-of-mass mode in this new basis of bath operators, enabling one to express all quantities in terms of scattering operators. Finally, we specify this approach to optomechanical systems, assuming a strongly-driven cavity and the Lamb-Dicke regime for the cm operator in Sec.~\ref{sec:optomech}. 
\subsection{Solution of the scattering equations for the free field}\label{sec:freefield}
\begin{figure}
\includegraphics[width=0.9\linewidth]{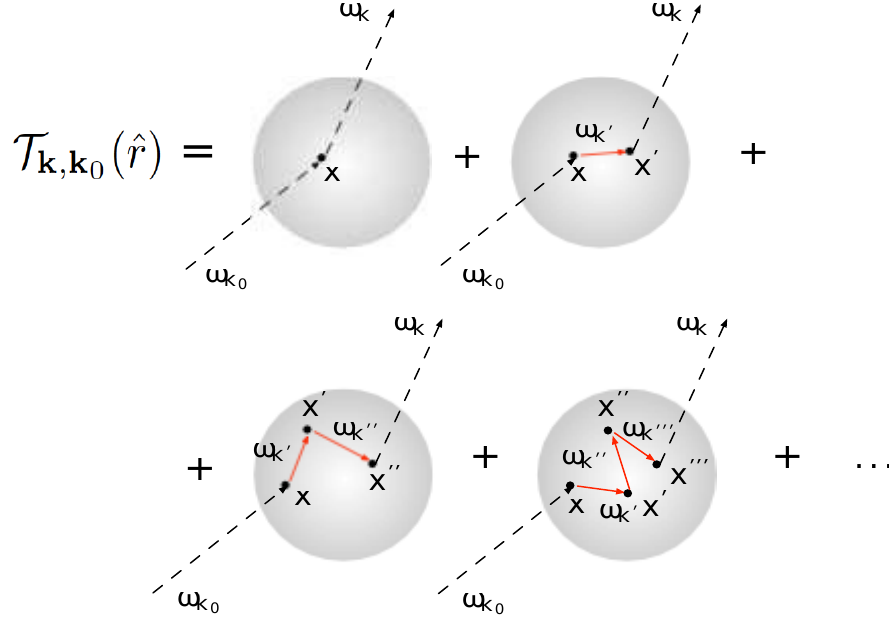}
\caption{Graphical illustration of the transition matrix $\mathcal{T}_{\bk, \bk_0}(\br)$ as an infinite series in the light-matter interaction. The first term denotes the direct interaction between the dielectric and light, the second one a process, where a photon is virtually absorbed and reemitted, the third one a process, where two intermediate photons are involved, etc.}\label{Fig:scattering}
\end{figure}

The equations of motion of the electromagnetic field in the Heisenberg picture are determined and connected to the Lippmann-Schwinger equation. We are only interested in the homogeneous part of the electromagnetic field $\Eop_{\rm B}(\bx,t)$ given by Eq.~\eqref{eq:freefield} and assume that no inhomogeneity (\ie, cavity) is present, leaving us with a system fully described by $\hat{H}_{\rm B}$, Eq.~\eqref{eq:bath}. We keep the center-of-mass operator $\br$ in the equations of motion, but neglect its action for now assuming $M\rightarrow \infty$.
The Heisenberg eqs. of motion can thus be determined as
\be \label{eq:adot}
\begin{split}
\dot{a}_{\bold{k}}(t)=-\im\omega_{\bold{k}}\hat{a}_{\bold{k}}(t)\!-\im\sqrt{\frac{\epsilon_c^2\epsilon_0\omega_{\bold{k}}}{2 (2\pi)^{3}}}\int_{V(\br)} d\bx  \Eop_{\rm B}(\bold{x},t) e^{i\bold{k}\bold{x}}. 
\end{split}
\ee
Let us first define $\Eop_{\rm B}=\Eop_{\rm B}^{(+)}+\Eop_{\rm B}^{(-)}$, where $\Eop_{\rm B}^{(-)}=\Eop_{\rm B}^{(+){\dagger}}$ with
\be
\Eop^{(+)}_{\rm B}(\bx,t)=\frac{\im}{(2\pi)^{3/2}}\int d \bk\sqrt{\frac{\omega_{\bold{k}}}{2\epsilon_0}}e^{-\im \bold{k} \bold{ x}}\aop_{\bold{k}}(t)
\ee
and the incoming field is given by
\be\label{eq:ein}
\Eop_{\rm B, \rm in}^{(+)}(\bx,t)=\frac{\im}{(2\pi)^{3/2}}\int d \bk\sqrt{\frac{\omega_{\bold{k}}}{2\epsilon_0}}e^{-\im \bold{k} \bold{ x}}\aop_{\bold{k}}(0)e^{-\im \omega_{\bk}t}.
\ee
With these definitions at hand, we can close the set of equations given by Eq.~\eqref{eq:adot} by carrying out the following steps: we formally integrate Eq.~\eqref{eq:adot} over time, multiply both sides by $\im \sqrt{\omega_{\bk}/2\epsilon_0(2\pi)^3}e^{-\im \bold{k}\bold{x}}$, and take the integration over $\bk$ to obtain
\be
\begin{split}
\Eop^{(+)}_{\rm B}(\bold{x},\br,t)=&\Eop^{(+)}_{\rm B,\rm in}(\bold{x},t) \\
&+ \int_{V(\br)} d\bx' \!\int d\bk \frac{\epsilon_c\omega_{\bold{k}}}{2(2\pi)^{3}} e^{\im\bold{k}(\bold{x}'-\bold{x})} e^{-\im \omega_k t}\\
& \times \Big[\Eop'^{(+)}_{\rm B}(\bold{x}',  t)\int_0^t d\tau e^{-\im(\omega_0-\omega_k) \tau}\\
& \; \; \; \; \;\;+\Eop'^{(-)}_{\rm B}(\bold{x}',  t)\int_0^td\tau e^{\im(\omega_0+ \omega_k)\tau}\Big]. 
\end{split}
\label{effE}
\ee
Note that we assume the spectral distribution of the electromagnetic field to be peaked at a certain frequency $\omega_0$ and we thus have introduced the slowly-varying field $\Eop'_{\rm B}(\bold{x},t)$ here, $\Eop'^{\pm}_{\rm B}(\bold{x},t)=e^{\pm \im \omega_0t}\Eop^{\pm}_{\rm B}(\bold{x})$. This justifies the assumption that  $\Eop'^{\pm}_{\rm B}(\bold{x},t)$ remains constant on the time scales of the system's evolution which allows one to take it out of the integration in Eq.~\eqref{effE}.
Integrating $d \bk$ in Eq.~\eqref{effE} yields a function that decays quickly in $\tau$. This allows for an extension of the upper integration boundary $t$ to $\infty$ (Markov approximation) and hence yields
\be
\begin{split}
\Eop_{\rm B}^{(+)}(\bold{x},\br,t)=&\Eop^{(+)}_{\rm B, \rm in}(\bold{x},t)+\frac{\epsilon_c }{2 (2\pi)^3}\int_{V(\br)} d\bx' \int d\bk \\
&\times e^{-\im \bk (\bx-\bx')}\frac{\omega_{\bk}}{\omega_{\bk}-\omega_0+\im \gamma}
\Eop^{(+)}_{\rm B}(\bold{x}', t)\label{eq:elfield},
\end{split}
\ee
where the limit $\gamma\rightarrow 0^+$ is understood.
Taking the inverse transformation, the field operators can be written as
\be\label{eq:fieldop}
\begin{split}
\hat{a}_{\bk}(t)=&\hat{a}_{\bk}(0)e^{-\im\omega_{\bk}t}
+\frac{\epsilon_c}{2}\int_{V(\br)}d\bx'\int d\bk'\times \\
&\times e^{\im(\bk-\bk')\bx'}\frac{\sqrt{\omega_{\bk'}\omega_{\bk}}}{\omega_{\bk'}-\omega_0+\im\gamma}\hat{a}_{\bk'}(t).
\end{split}
\ee
This equation for the operators of the electromagnetic field resembles the Lippmann-Schwinger equation~\cite{goldbergerbook, sakuraibook}.  In order to connect the two descriptions, enabling one to employ solutions known from scattering theory in our approach, we proceed in the same way defining
\be
\begin{split}
\mathcal{V}_{\bk, \bk'}(\br)=\frac{\epsilon_c}{2}\int_{V(\br)}d\bx'\sqrt{\omega_{\bk'}\omega_{\bk}}e^{\im(\bk-\bk')\bx'}
\end{split}
\ee
as the matrix elements of the operator describing the scattering interaction. In analogy we define the transition matrix $\mathcal{T}_{\bk, \bk'}(\br)$, given by
\be
\begin{split}
\mathcal{T}_{\bk, \bk'}(\br)=&\int d\bk''\mathcal{V}_{\bk, \bk''}(\br)\times\\
&\times\left(\delta(\bk'-\bk'')+\frac{\mathcal{T}_{\bk''\bk'}(\br)}{\omega_{\bk''}-\omega_{\bk}+\im\gamma}\right).\label{eq:trans}
\end{split}
\ee
Note that both $\mathcal{V}_{\bk, \bk'}(\br)$ and $\mathcal{T}_{\bk, \bk'}(\br)$ are operators for the center-of-mass degree of freedom but not for the photonic ones. That is, if we fix $\br$, neglecting the object's motion, $\mathcal{V}_{\bk, \bk'}(\br)$ and $\mathcal{T}_{\bk, \bk'}(\br)$ are simply numbers without any operator-character. By iteration, the transition matrix describes scattering processes to all orders of perturbation theory, as illustrated in Fig.~\ref{Fig:scattering} (an explicit formula for the expansion of $\mathcal{T}_{\bk, \bk'}(\br)$ in terms of $\mathcal{V}_{\bk, \bk'}(\br)$ can be found in standard textbooks~\cite{sakuraibook}).

Subsequently, Eq.~\eqref{eq:trans} enables one to rewrite the total time evolution of the operators, Eq.~\eqref{eq:fieldop}, as
\be
\begin{split}
\hat{a}_{\bk}(t)=&\int d\bk'e^{-\im\omega_{\bk'}t}\times\\
&\times\left(\delta(\bk-\bk')
+\frac{\mathcal{T}_{\bk, \bk'}(\br)}{\omega_{\bk'}-\omega_0+\im\gamma}\right)\hat{a}_{\bk'}(0).\label{eq:evak}
\end{split}
\ee
This expression is equivalent to the classical field equations given by Eq.~\eqref{elfield_cl}. For its solution we can thus rely on the variety of methods that has been developed during the past decades described in Sec.~\ref{classical}. 

A useful relation that will be used to simplify the computation of transition amplitudes for spheres in Sec.~\ref{sec:spheres}, is the optical theorem connecting the scattering amplitude in forward-direction to the scattering in all other directions~\cite{goldbergerbook}:
 \be
 \Im[\mathcal{T}_{\bk, \bk'}(\br)]=-\pi\int d\bk'|\mathcal{T}_{\bk, \bk'}(\br)|^2\delta(\omega_{\bk}-\omega_{\bk'}).\label{eq:opt_th}
 \ee

Before we continue the analysis, to ease the notation, it is useful to define the space of mode functions in which the matrices $\mathcal{T}_{\bk, \bk'}(\br)$ and $\mathcal{V}_{\bk, \bk'}(\br)$ act, and consider them as operators, \ie,  $\mathcal{V}_{\bk, \bk'}(\br)=\langle\bk|\hat V(r)|\bk'\rangle$, where $|\bk\rangle$ are the basis vectors of such a space. They can be viewed as mode functions with momentum $\bk$. As we describe scattering out of the cavity mode in this article, let us now define the transition amplitudes for mode shapes different from plane waves and express them in the basis $|\bk\rangle$.
For the Born approximation of scattering theory~\footnote{Note, that we refer to the Born approximation of scattering theory here, where the action of the scattering event on the electromagnetic field is neglected in the lowest-order approximation. Thus, no multi-scattering events are accounted for. This is different from the Born-Markov approximation often taken in the derivation of master equations, which assumes the separability of the density matrices of the system and the bath.}, consisting in setting $\hat{T}(\br)\approx\hat{V}(\br)$, we obtain
\be
\mathcal{V}_{\bk, \rm c}(\br)=\int d\bk'\langle  \bk|\hat{V}(\br)|\bk'\rangle\langle\bk'|\rm c\rangle,
\ee
where $|\rm c\rangle$ describes the mode function of the cavity, which can be written as $\langle \bx|{\mathrm c}\rangle=f(\bx)/\sqrt{  V_c}$ in position-representation, where $f(\bx)$ is assumed to be real.
Evaluating this expression yields
\be
\begin{split}
\mathcal{V}_{ \bk, \mathrm c}(\br)=&\sqrt{\frac{\epsilon_c^2\omega_{\bk}\omega_0}{ 4 V_c}}\int_{V(\br)}d\bx f(\bx)e^{\im \bk\bx},\label{eq:expV}
\end{split}
\ee
where we have used that the distribution of $\omega_{\bk}$ is peaked around $\omega_0$, allowing for the substitution $\omega_{\bk}\approx\omega_0$.

\subsection{General master equation for the cavity and the center-of-mass mode}\label{sec:cavcm}

Within the Born-Markov approximation, the master equation describing the system's full dynamics is given by
\be
 \dot{\rho}_{\mathrm S} (t) \approx - \int_0^\infty  \tr_{\mathrm B} [\hat{H}_{\mathrm BS}^{\mathrm I}(t), [\hat{H}_{\mathrm BS}^{\mathrm I}(t-\tau),\hat{\rho}_{\mathrm S}(t) \otimes \hat{\rho}_{\mathrm B}]] d \tau.\label{eq:master}
\ee
The Born-Markov approximation consists in the following assumptions: the density matrices of the system and the environment are considered to be separable, $\hat{\rho}_{\mathrm {tot}}=\hat{\rho}_{\rm S}\otimes\hat{\rho}_{\rm B}$, and correlations between bath operators are taken to decay quickly. Furthermore, the bath is assumed to remain unchanged during the interaction with the system, $\hat{\rho}_{\rm B}(t)\approx\hat{\rho}_{\rm B}(0)$. This is valid given that the bath is very large and the effect of the interaction with the system can be neglected. 

Moreover, for typical quantum-optical systems interactions between different bath operators $\aop_{\bk}, \adop_{\bk'}$, Eq.~\eqref{eq:bathint}, are negligible, \ie,  $\hat{W}(\br)\approx 0$, such that 
\be
\hat{H}_0\approx\hat{H}_{\rm S}+\int d\bk \omega_{\bk}\adop_{\bk}\aop_{\bk}.
\ee 
While these approximations are typically fulfilled for point-particles, difficulties are encountered when extending the method to larger objects. 
It is in particular the negligence of interactions between different bath operators that is no longer justified. 
More specifically, we realize that contributions, where interactions among bath operators are taken into account to different orders, scale as $\propto (R/\lambda)^{2n}$. Here, $R$ denotes the dimensions of the object, $\lambda$ the wavelength of the inhomogeneous light mode, and $n$ the $n$th order of the multiple scattering process.\\
Consequently, an approach where these interactions are accounted for is necessary. This is done by including the correlations between the bath operators described by $\hat{W}(\br)$ in the Hamiltonian that is used to transform to the interaction picture, 
\be
\hat{H}_{\rm BS}^{\rm I}(t)=e^{\im \hat{H}_0t}\hat{H}_{\rm BS}(0)e^{-\im \hat{H}_0t}.
\ee
To find a solution, we connect this approach to the description within scattering theory given in Sec.~\ref{sec:freefield}. Based on this analysis, we can develop a master equation that accounts for interactions among different bath modes.
As an example we now discuss the first term of the master equation, where all operators are in front of the density matrix, in more detail: 
\be
\begin{split}
\dot{\rho}_{\rm S}=&-\adop_0\aop_0\int_0^{\infty}d\tau\int d\bk e^{\im\omega_0\tau} \langle \Omega |\mathcal{F}(t,\br)\aop_{\bk}^{\dagger}(0)|\Omega\rangle\times\\
&\times\langle\Omega|\aop_{\bk}(0)\mathcal{F}(t-\tau,\br)|\Omega\rangle \hat{\rho}_{\rm S}+...\label{eq:master1}
\end{split}
\ee 
Here, $|\Omega\rangle$ denotes the vacuum state and we have defined
\be
\mathcal{F}(t,\br)=\sqrt{\frac{\epsilon_c^2\omega_0\epsilon_0}{4V_c}}\int_{V(\br)}d\bx f(\bx)\Eop_{\rm B}(\bx,t),
\ee
where counter-rotating terms have been neglected.
Let us now connect Eq.~\eqref{eq:master1} to the description in terms of mode functions in the scattering picture. First, we shift the time dependance of $\mathcal{F}(t,\br)$ to the operators by
\be
\begin{split}
\langle \Omega |\mathcal{F}(t,\br)\aop_{\bk}^{\dagger}(0)|\Omega\rangle
=\langle \Omega| \mathcal{F}(0,\br)\aop_{\bk}^{\dagger}(-t)|\Omega\rangle,\label{eq:fop}
\end{split}
\ee
where the invariance of the vacuum state under time evolution has been used. In order to make the procedure more transparent,  as a first step, only the $0$th order Born approximation of scattering theory is identified. Subsequently, the treatment is extended to a description of all orders.
The lowest order of the Born series gives $\aop_{\bk}(t)\approx\aop_{\bk}(0)\exp(-\im\omega_{\bk}t)$, so that we need to evaluate
\be
\begin{split}
\langle \Omega| \mathcal{F}(0,\br)\aop_{\bk}^{\dagger}(0)|\Omega\rangle=\sqrt{\frac{\epsilon_c^2\omega_{\bk}\omega_0}{4 V_c}}\int_{V(\br)}d\bx f(\bx)e^{-\im \bk\bx}.
\end{split}
\ee
Recalling the definition of the expectation value $\mathcal{V}_{\bk, \rm c}(\br)$ in the scattering picture, Eq.~\eqref{eq:expV}, we identify
\be
\langle \Omega| \mathcal{F}(0,\br)\aop_{\bk}^{\dagger}(0)|\Omega\rangle=\mathcal{V}^*_{\rm c, \bk}(\br).
\ee
The same procedure can now be applied without taking the Born approximation and considering the full transition matrix, by plugging Eq.~\eqref{eq:evak} into Eq.~\eqref{eq:fop}, which yields
\be
\begin{split}
&\langle \Omega|\mathcal{F}(0,\br)\aop_{\bk}^{\dagger}(-t)|\Omega\rangle\\
&=\int d\bk'\mathcal{V}^*_{\rm c, \bk'}(\br)\left(\delta(\bk-\bk')+\frac{\mathcal{T}^*_{\bk'\bk}(\br)}{\omega_{\bk}-\omega_0-\im\gamma}\right)e^{-\im\omega_{\bk}t}\\
&=e^{-\im\omega_{\bk}t}\mathcal{T}^*_{ \rm c, \bk}(\br),
\end{split}
\ee
where Eq.~\eqref{eq:trans} has been used.

All other terms of the master equation can be determined in full analogy yielding
\begin{widetext}
\be
\dot{\rho}_{\rm S}=\im[\hat{\rho}_{\rm S}, \hat{H}_{S}+\hat{H}_{\rm rn}]+\int d\bk  \delta(\omega_{\bk}-\omega_0)\left(2\mathcal{T}_{\bk, \rm c}(\br)\aop_0\hat{\rho}_{\rm S} \adop_0\mathcal{T}^{*}_{\rm c,  \bk}(\br)-\left[| \mathcal{T}_{ \bk, \rm c}(\br)|^2\adop_0\aop_0, \hat{\rho}_{\rm S}\right]_{+}\right)\label{eq:master_dens_tot}, 
\ee
\end{widetext}
where $\hat{H}_{S}$ is the system Hamiltonian given by Eq.~\eqref{eq:Hsys} and $\hat{H}_{\rm rn}$ the renormalization
\be
\hat{H}_{\rm rn}=\adop_0\aop_0\int d\bk |\mathcal{T}_{\bk,\rm c}(\br)|^2\mathcal{P}\frac{1}{\omega_{\bk}-\omega_0}, 
\ee
where $\mathcal{P}$ denotes Cauchy's principal value. Note that a similar master equation for the cm degree of freedom has been discussed in the context of scattering of air molecules~\cite{Diosi95, hornberger03, Adler06, hornberger06, hornberger07}.
\subsection{Master equation for the optomechanical setup}\label{sec:optomech}
In this section, we adapt the general master equation Eq.~\eqref{eq:master_dens_tot} to the specific optomechanical setup we are interested in. Therefore, we take the following approximations:
\begin{enumerate}
\item The inhomogeneous mode is assumed to be a strongly-driven cavity effecting large cavity occupation numbers $n_{\rm phot}=|\alpha|^2$, such that $\alpha\gg \langle\aop_0\rangle$. This enables one to neglect certain terms in the master equation.

\item We assume the Lamb-Dicke regime: the dielectric is positioned close to the maximal slope of the standing wave in the cavity and close to the minimum of the harmonic trapping potential of the optical tweezers. The motion around its cm position is considered to be small, such that the Lamb-Dicke parameter $\eta=k \Delta \br\ll 1$ (with $\Delta \br=\sqrt{\langle \br^2 \rangle-\langle\br\rangle^2}$), facilitating an expansion of the transition operator matrix elements $\mathcal{T}_{\bk, \rm c}(\br)$ in $\bk \br$.
\end{enumerate}
Displacing the cavity operator by $\alpha$ such that $\aop_0\rightarrow \aop_0'+\alpha$ and expanding the transition operator to second order, $\mathcal{T}_{ \bk, \rm c}(\br)\approx \mathcal{T}_{ \bk, \rm c}(0)+\mathcal{T}_{ \bk, \rm c}'(\br)|_{\br=0} \br+ \mathcal{T}_{\bk, \rm c}''(\br)|_{\br=0} \br^2$ leads to a master equation, where we take into account terms that are at most of quadratic order in the cavity operators $\aop_0, \adop_0$ and the cm operators $\br=x_0(\bop+\bdop)$. Here, $\mathcal{T}_{\bk, \rm c}^n(\br)=\partial^n \mathcal{T}_{\bk, \rm c}(\br)/\partial \br ^n$ denotes the $n$th partial derivative and $x_0=\sqrt{1/2M\omega_t}$ the zero-point motion of the center-of mass mode. 

In the following we give an interpretation of the different contributions to the master equation and indicate which terms yield a renormalization to the Hamiltonian, can be neglected, or describe decoherence. We describe these terms in decreasing order in $\alpha$.

\subsubsection{Contributions $\propto |\alpha|^2$}
The largest contribution to the master equation are terms $\propto |\alpha|^2 |\mathcal{T}_{ \bk, \rm c}(0)|^2$. As they do not contain operator-character they cancel.

The next order in the Lamb-Dicke parameter $\eta$ is given by terms $\propto \br$, which can be shown to vanish using Hilbert transforms and the analytic property of the function $\mathcal{T}'_{ \bk, \rm c}(\br)|_{\br=0} \mathcal{T}_{ \rm c, \bk}^{*}(0)$, see App.~\ref{master_opto} for an explicit analysis.

The only contributing terms are $\propto \br^2$ and describe a renormalization of the trapping frequency of the dielectric provided by the optical tweezers and decoherence of the cm operator. The renormalization of the trapping frequency  $\tilde\omega_t=\omega_t+\Delta^{\rm M}$ can be simplified exploiting the analytic properties of the functions and is given by
\be\label{eq:delta}
\Delta^{\rm M}=|\alpha|^2x_0^2\int d\bk \mathcal{P}\frac{1}{\omega_{\bk}-\omega_0}\left[\mathcal{T}'_{ \bk, \rm c}(\br) \mathcal{T}_{ \rm c, \bk}'^{*}(\br)\right]_{\br=0}.
\ee 
The decoherence of the mechanical motion is described by
\be
\begin{split}
\mathcal{L}^{\rm M}[\hat{\rho}_{\rm S}]&=\Gamma \left(2(\bop+\bdop)\hat{\rho}_{\rm S}(\bop+\bdop)-\left[(\bop+\bdop)^2, \hat{\rho}_{\rm S}\right]_+\right)
\label{eq:deccms}
\end{split}
\ee
with
\be
\Gamma=|\alpha|^2x_0^2\int d\bk\delta(\omega_{\bk}- \omega_0)\left[\mathcal{T}'_{  \bk, \rm c}(\br) \mathcal{T}_{  \rm c, \bk}'^{*}(\br)\right]_{\br=0}.\label{eq:decmech}
\ee

The decoherence of the cm thus depends on the form of the transition amplitudes with respect to the cm position. The physical process underlying this effect is recoil heating via photon scattering.
\subsubsection{Contributions $\propto \alpha$}
Also for contributions $\propto\alpha^*\aop_0 |\mathcal{T}_{ \bk, \rm c}(0)|^2$, the analyticity of the transition operator can be exploited. Applying a Hilbert transformation, we can show that these contributions cancel, see App.~\ref{master_opto} for a more detailed analysis. 

Terms $\propto\alpha^* \aop_0 \br$ effect both a coherent and an incoherent contribution. The incoherent part describes decoherence of the mechanical and the light degree of freedom and can be shown to be negligible as demonstrated in App.~\ref{master_opto}.
In contrast, the coherent contribution yields a non-negligible renormalization of the optomechanical coupling $\tilde g=g+g_{\rm rn}$ defined by
\be\label{eq:gopto}
g_{\rm rn}=\alpha^*x_0\int d\bk\mathcal{P}\frac{1}{\omega_{\bk}-\omega_0}\mathcal{T}_{\bk, \rm c}(0) \mathcal{T}'^{*}_{ \rm c, \bk}(\br)|_{\br=0}.
\ee
Furthermore, terms $\propto \br^2$ describe decoherence of both the mechanical mode and the light mode. Comparing to Eq.~\eqref{eq:decmech} for the cm mode, these contributions are suppressed by $1/\alpha$ and can thus be neglected. Also for the cavity mode, these terms are negligible, given that $\eta^2\ll1/\alpha$.

\subsubsection{Contributions $\propto \adop_0\aop_0$}
Terms $\propto |\mathcal{T}_{\bk, \rm c}(0)|^2$ yield both a coherent and an incoherent contribution describing a renormalization of the resonance frequency of the cavity and a part describing the cavity's decay:
\be
\begin{split}
\mathcal{L}^{\rm L}[\rhos]=&\kappa\left(2\aop_0\rhos\adop_0-[\adop_0\aop_0,\rhos]_+\right)\label{eq:deccav}
\end{split}
\ee
with 
\be
\kappa=\int d\bk \delta(\omega_{\bk}-\omega_0) |\mathcal{T}_{ \bk, \rm c}(0)|^2.\label{eq:kappa}
\ee
The renormalization of the cavity's resonance frequency $\tilde\omega_0=\omega_0+\Delta^{\rm L}$ is defined by
\be
\Delta^{\rm L}=\int d\bk \mathcal{P}\frac{1}{\omega_{\bk}-\omega_0} |\mathcal{T}_{\rm c, \bk}(0)|^2.\label{eq:rncav}
\ee
These are the only non-vanishing contributions as terms $\propto \br$ are suppressed by the Lamb-Dicke parameter $\eta$ and terms $\propto \br^2$ even by  $\eta^2$ compared to Eqs.~\eqref{eq:deccav}, ~\eqref{eq:rncav}.

\subsubsection{Final master equation}
To summarize, we identify the contributions to the final master equation:
\be
\begin{split}
\dot{\rho}_{\rm S}=&\im[\hat{\rho}_{\rm S},\hat{H}_S+\hat{H}_{\rm rn}]+\mathcal{L}^{\rm M}[\hat{\rho}_{\rm S}]+\mathcal{L}^{\rm L}[\hat{\rho}_{\rm S}].\label{eq:mastertot}
\end{split}
\ee
They can be grouped as follows:
\begin{enumerate}
\item  Contributions of Hamiltonian-type,
\be
\begin{split}
\hat{H}_S+\hat{H}_{\rm rn}=&-\delta \adop_0\aop_0+\tilde\omega_t\bdop \bop+\tilde g(\aop_0+\adop_0)(\bop+\bdop),\label{eq:hsysopt}
\end{split}
\ee
where the frequencies and couplings stemming from the system's Hamiltonian
$\hat{H}_{\rm S}$, given by Eq.~\eqref{eq:Hsys}, are renormalized by
\be
\hat{H}_{\rm rn}=\Delta^{\rm M}\bdop\bop+\Delta^{\rm L}\adop_0\aop_0+g_{\rm rn}(\aop_0+\adop_0)(\bop+\bdop).
\ee 
The corresponding renormalizations are defined by Eqs.~\eqref{eq:delta},~\eqref{eq:gopto},~\eqref{eq:rncav}.
Note that the Hamiltonian of Eq.~\eqref{eq:hsysopt} has been transformed to a frame rotating at the laser frequency $\omega_L$, where $\delta$ now denotes its detuning from the cavity resonance frequency $\tilde \omega_0$.
\item The recoil heating via photon scattering of the cm mode yields $\mathcal{L}^{\rm M}[\hat{\rho}_{\rm S}]$, given by Eq.~\eqref{eq:deccms}.
\item The decay of the cavity mode due to the presence of the object yielding $\mathcal{L}^{\rm L}[\hat{\rho}_{\rm S}]$, is described by Eq.~\eqref{eq:deccav}.

\end{enumerate}
Consequently, all frequencies, couplings, and decay rates are renormalized taking into account all terms beyond the first Born approximation of scattering theory. This enables one to use exact solutions if available, or in general to truncate the perturbation series in a controlled way.

While this master equation only contains the time evolution of the cavity and the cm operators, information about the scattered fields can be obtained by applying the quantum regression theorem. The scattered light is directly accessible in experiments and can, \eg, be used to monitor the cooling of the mechanical motion~\cite{cirac93}. To complement the analysis given here, we show how to derive the scattered fields directly in App.~\ref{sec:wigner} within an approach similar to Wigner-Weisskopf, but accounting for interaction processes between the bath modes.

\section{Cooling}\label{sec:cooling}

Before applying the master equation discussed in the previous section to the particular case of cavity optomechanics with optically levitating spheres, we provide a general description of optomechanical cooling and the minimal phonon number attainable with master equations of at most quadratic order in the operators of the mechanical and the cavity mode. Cooling is in general a vital ingredient in any attempt to demonstrate quantum-mechanical behavior. Most descriptions make certain approximations to ease calculations, \eg, the sideband regime is commonly employed in optomechanical setups~\cite{Marquardt07, WilsonRae07}. 
Here, the system will be treated in the most general way not relying on any approximations (as some of them might not be fulfilled for larger objects). The master equations we are interested in, are typically of the form given by Eq.~\eqref{eq:mastertot}.

\begin{figure}
  \includegraphics[width=.48\linewidth]{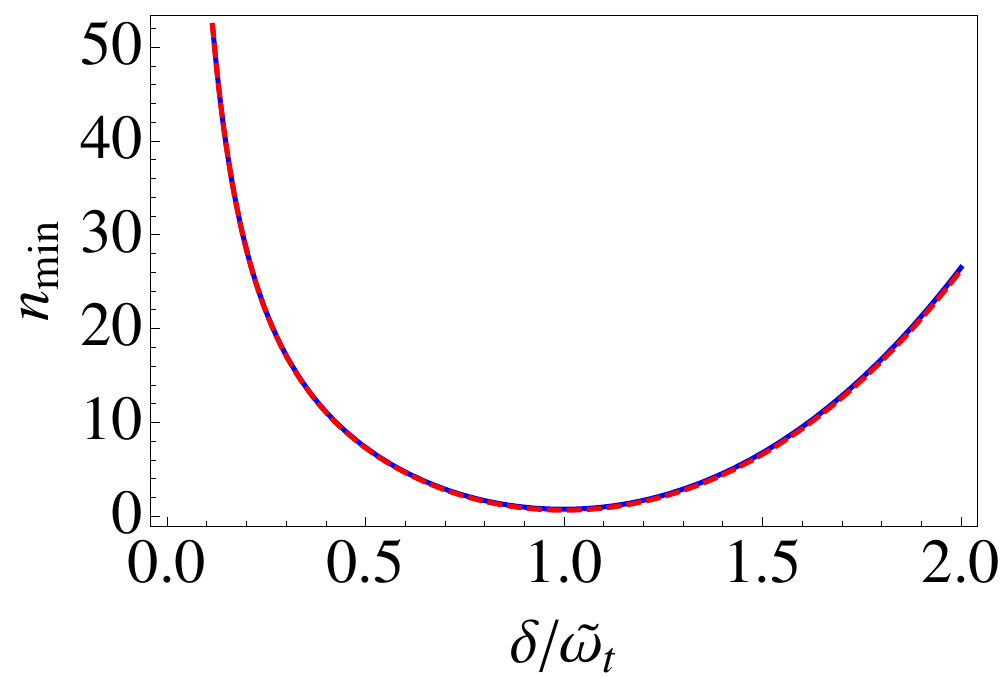}
 \includegraphics[width=.48\linewidth]{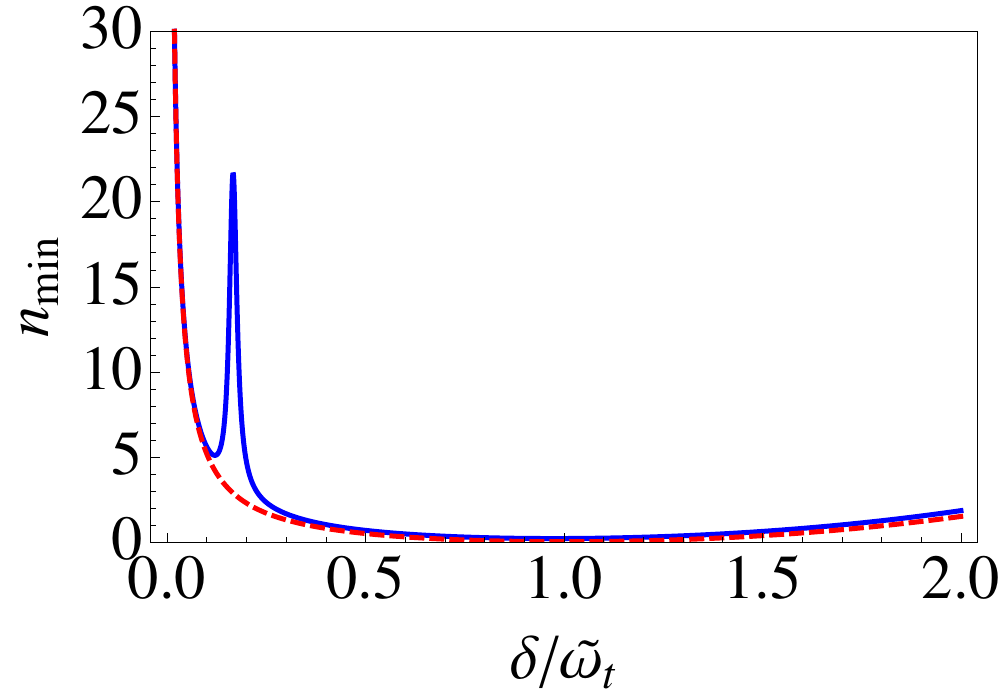}
 \includegraphics[width=.48\linewidth]{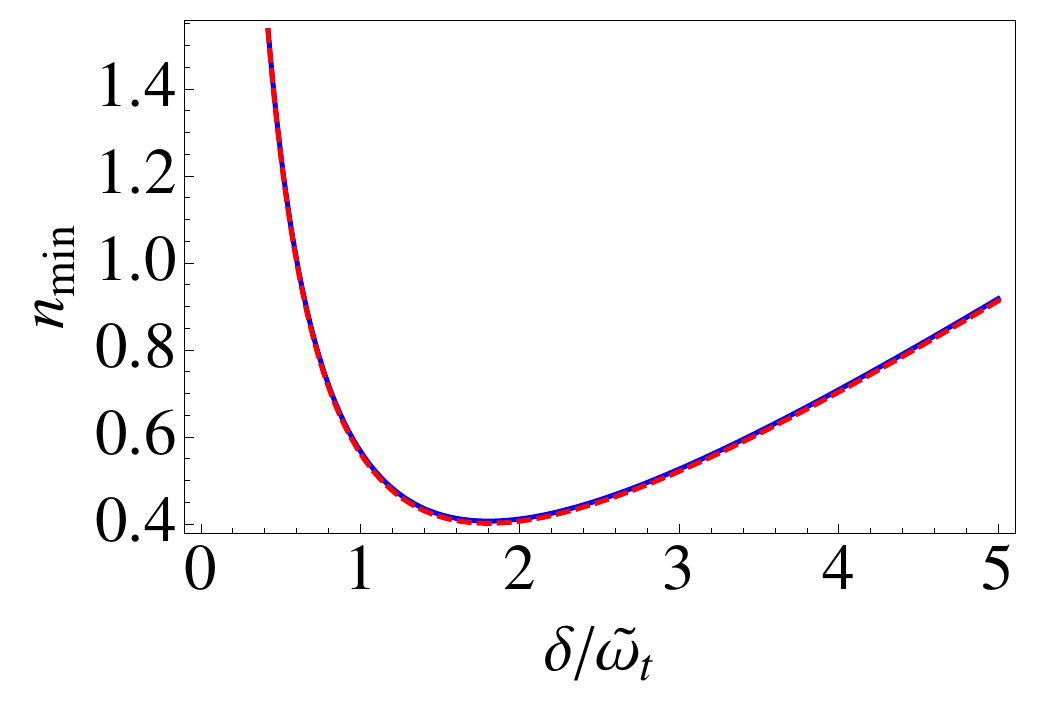}
 \includegraphics[width=.48\linewidth]{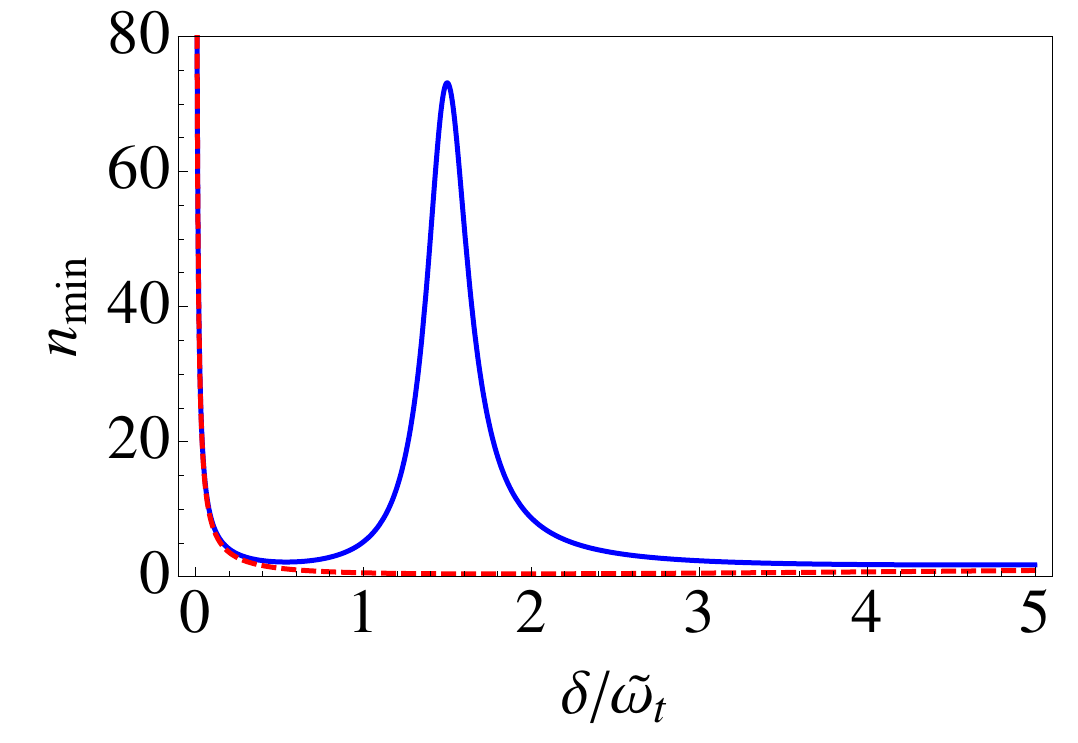}
\caption {Minimal phonon numbers for different detunings comparing the exact result (blue straight line) to the one with the cavity mode eliminated (red dashed line). Differences between the two solutions evolve as $\tilde{g}$ is increased. Only the regions, where a steady state is attainable have been plotted, they all lie within the red-sideband regime. For blue detuning or an optomechanical coupling which is too strong, the system is heated and no steady state can be obtained.
\textit{Upper left pannel:} Sideband-resolved regime with weak coupling, $\kappa=0.3~\tilde{\omega}_{\rm t}, \Gamma=0.03~\tilde{\omega}_{\rm t}, \tilde{g}=0.07~\tilde{\omega}_{\rm t}$, \textit{Upper right pannel:} $\kappa=0.3~\tilde{\omega}_{\rm t}, \Gamma=0.03~\tilde{\omega}_{\rm t}, \tilde{g}=0.3~\tilde{\omega}_{\rm t}$, \textit{Lower left pannel:} Bad-cavity limit for weak coupling $\kappa=3~\tilde{\omega}_{\rm t}, \Gamma=0, \tilde{g}=0.1~\tilde{\omega}_{\rm t}$,\textit{Lower right pannel:} Bad-cavity limit for strong coupling, $\kappa=3~\tilde{\omega}_{\rm t}, \Gamma=0, \tilde{g}=0.864~\tilde{\omega}_{\rm t}$}
\label{Fig:nmin}
\end{figure}

The mean phonon number $\langle \bdop \bop \rangle$ is coupled to all other expectation values of combinations of the operators $\aop_0, \adop_0, \bop$, and $\bdop$, yielding the Eqs. of motion in matrix form
\be
\dot{\bold{v}}=M\bold{v}+\bold{c},\label{eq:master_vec}
\ee
where
\be
\bold{v}=( \langle \bdop \bop\rangle, \langle \aop_0 \bdop \rangle,\langle \adop_0 \adop_0 \rangle, \langle \adop_0 \bop \rangle,....)^T,
\ee
$M$ denotes the interaction matrix, and $\bold{c}$ a constant vector. The master equation Eq.~\eqref{eq:mastertot} keeps the Gaussian character for an initially Gaussian state. Subsequently, also the system of equations Eq.~\eqref{eq:master_vec} is Gaussian in the operators $\aop_0, \adop_0...$ and linear in the expectation values $\langle\bdop \bop\rangle$. Since Eq.~\eqref{eq:master_vec} represents a closed system of equations, they can be solved as
\be
\bold{v}=e^{M t}\bold{v}(0)+\left(e^{M t}-1\right)M^{-1}\bold{c}, \label{general_v}
\ee  
yielding 
\be
\bold{v}=-M^{-1}\bold{c}
\ee  
for the steady state. The steady state phonon number can be extracted from this quantity as
\be
\begin{split}
\bar{n}=\rm{lim}_{t\rightarrow \infty}\langle \bdop \bop\rangle=\frac{A_1-A_2+A_3}{A_4}.\label{eq:nsteady}
\end{split}
\ee
with $A_1=32 \tilde g^4 \delta  [4 \delta ^2 \kappa +\kappa ^3+16 \delta  (\Gamma -\kappa ) \tilde\omega_{\rm t} +8 \kappa \tilde\omega_{\rm t} ^2]$, $A_2=-\Gamma  [4 \delta ^2+\kappa ^2] \tilde\omega_{\rm t} [16 \delta ^4+8 \delta ^2 \left(\kappa ^2-4 \tilde\omega_{\rm t} ^2\right)+\left(\kappa ^2+4 \tilde\omega_{\rm t} ^2\right)^2]$, $A_3=4 \tilde g^2 \{-\kappa  (4 \delta ^2+\kappa ^2) [\kappa ^2+4 (\delta -\tilde{\omega}_{\rm t} )^2] \tilde\omega_{\rm t} +2 \Gamma  \delta  [32 \tilde\omega_{\rm t} ^4+(4 \delta ^2+\kappa ^2)^2+4 (-20 \delta ^2+3 \kappa ^2) \tilde\omega_{\rm t}^2]\}$, and $A_4=64 \tilde g^2 \delta  \kappa  \tilde\omega_{\rm t} [16 \tilde g^2 \delta -\left(4 \delta ^2+\kappa ^2\right) \tilde\omega_{\rm t}]$.
This solution might contain unphysical results. To verify that $\bar{n}$ is indeed a steady state of the system, the eigenvalues of M additionally have to fulfill $\Re[\rm{eig}[M]]\leq0$.
In general, all parameters are determined by the properties of the system, solely the detuning $\delta$ can be chosen. According to the definition of Eq.~\eqref{eq:hsysopt}, $\delta>0$ denotes red detuning and $\delta<0$ blue detuning. To obtain the optimal point for cooling, one consequently has to optimize $\bar{n}$ with respect to $\delta$. 
Let us now compare this exact solution to the one obtained after an adiabatic elimination of the cavity mode. Starting from Eq.~\eqref{eq:mastertot}, we eliminate the cavity mode assuming that its decay rate is much larger than the coupling between the mechanical degree of freedom and the light, $\kappa \gg \tilde{g}$. In this case it is justified to assume that the cavity is either empty or contains only one photon, therefore reducing the master equation to the one-excitation manifold, described by $\rho_{00}, \rho_{10}, \rho_{01}, \rho_{11} $. Due to the fast decay of the cavity mode described by $\kappa$, the change of all contributions involving an excitation is approximately zero, finally yielding an equation of motion for the empty cavity $\rho_{00}$. After carrying out a rotating wave approximation assuming $\tilde{\omega}_{\rm t} \gg |\tilde{g}^2 /(\kappa+\im (\delta \pm \tilde{\omega}_{\rm t}))|$, the final steady state phonon occupation is given by: 
\be
\bar{n}_{\rm adiab}=\frac{[4 \tilde{g}^2 \kappa+\Gamma (\kappa^2+4(\delta+\tilde{\omega}_{\rm t})^2)][\kappa^2+4(\delta -\tilde{\omega}_{\rm t})^2]}{64 \tilde{g}^2\delta \kappa \tilde{\omega}_{\rm t}}.\label{eq:nadiab}
\ee
To obtain the minimal occupation number, this equation needs to be optimized with respect to $\delta$. Comparing the adiabatically-eliminated solution to the exact one, it becomes clear that the approximation breaks down in the strong-coupling regime $ \tilde{g} \approx \tilde{\omega}_{\rm t}$, where the rotating-wave approximation is no longer valid,  see Fig.~\ref{Fig:nmin} for an illustration. Note that Eq.~\eqref{eq:nadiab} can be derived from Eq.~\eqref{eq:nsteady} taking the approximation $\kappa \gg \tilde{g}$. In case we choose the detuning $\delta=\tilde{\omega}_{\rm t}$ and $\tilde{\omega}_{\rm t}\gg\kappa$, we obtain the minimal occupation number in the sideband regime
\be
\bar{n}_{\rm sb}=\left(\frac{\kappa}{4\tilde{\omega}_{\rm t}}\right)^2+\frac{1}{4\mathcal{C}},
\ee
where $\mathcal{C}$ denotes the cooperativity, given by Eq.~\eqref{eq:coop}.  

\section{Cavity quantum-optomechanics with levitating spheres}\label{sec:impl_objects}\label{sec:spheres}
The theory developed in the previous sections provides a general framework to describe dielectrics interacting with one specific mode of the electromagnetic field. Here it will be used to describe cavity-optomechanics with levitating spheres~\cite{chang10, romeroisart10, Barker10a}. Due to the levitation, their mechanical degree of freedom (\ie, the motion of their cm) is prevented from clamping losses, which are the main source of decoherence in most optomechanical systems. Hence, they hold the promise for a variety of applications, ranging from the efficient implementation of protocols to realize non-Gaussian states~\cite{romeroisart2011a} to the preparation of large superpositions of their cm position~\cite{RomeroIsart11b, RomeroIsart11c}. In particular this enticing perspective has the potential to realize tests of quantum mechanics in an entirely new parameter regime. All of these results are only valid for nano-objects smaller than the optical wavelength, enabling one to neglect multiple-scattering effects. Although schemes to Doppler-cool dielectric spheres using Mie resonances have been discussed~\cite{barker10}, cavity-optomechanics with larger levitating objects has not been described before. In particular in the light of recent experiments on feedback-cooling of a microsphere~\cite{tongcang11}, a description of this new parameter regime is timely. 

\begin{figure}
\includegraphics[width=0.95\linewidth]{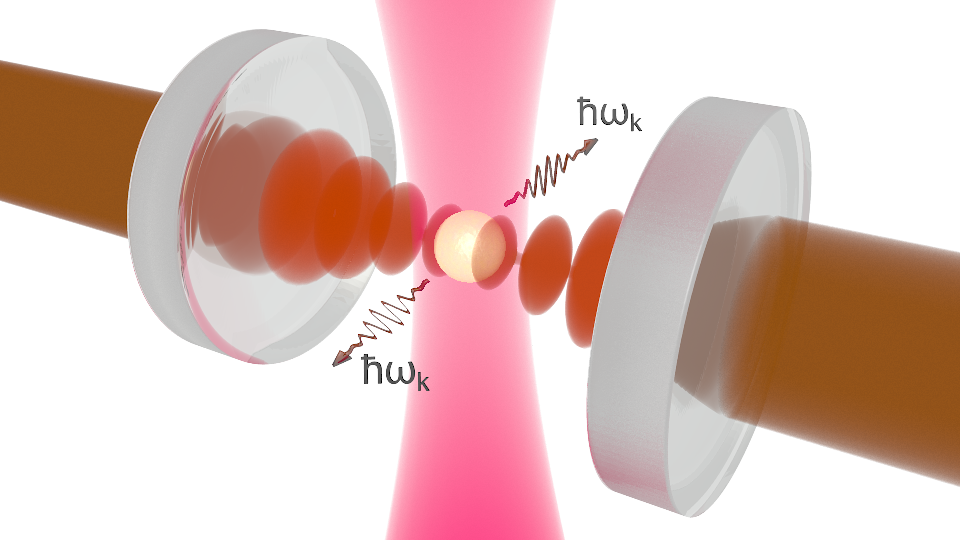}
\caption{Schematic representation of the setup: A dielectric sphere of a radius similar or larger than the cavity wave length is trapped by optical tweezers providing a trapping frequency $\tilde{\omega}_t$. It is placed inside an optical cavity, where a second laser is used to optically manipulate the dielectric's center-of-mass degree of freedom.}\label{Figure:setup1}
\end{figure}

We thus proceed to determine the optomechanical parameters and final occupation numbers for levitating dielectric spheres. The setup is similar to the one discussed in~\cite{romeroisart2011a}, but is extended to a description of spheres larger than the cavity wavelength. It is sketched in Fig.~\ref{Figure:setup1}: a classical light field, effected by a retro-reflected optical tweezer,  $\mathcal{E}_{\rm tw}(\bx,t)$,  creates a harmonic trap for the cm of the dielectric, (note that trapping via a strongly-populated cavity mode can be described in full analogy). 
Besides, a second cavity field $\Eop_{\rm cav}(\bx,t)$ is used to manipulate it, such that 
\be
\begin{split}
\Eop_{\rm S}(\bx,t)&=\Eop_{\rm cav}(\bx,t)\\
\mathcal{E}_{\rm S}(\bx,t)&=\mathcal{E}_{\rm tw}(\bx,t)+\mathcal{E}_{\rm cav}(\bx,t).
\end{split}
\ee
The optical tweezer used for the trapping is given by
\be
\mathcal{E}_{\rm tw}(\bx,t)=E_0 \frac{W_t}{W(y)}\exp\left(-\frac{x^2+z^2}{W(y)^2}\right)\label{eq:etw},
\ee
where $E_0=[P_t/(\epsilon_0 c \pi W_t^2)]^{1/2}$, $P_t$ is the laser power, $W_t$ is the laser beam waist, $W(y)=W_t[1+(y\lambda/(\pi W_t^2))^2]^{1/2}$ and we assume the beam to be aligned as sketched in Fig.~\ref{Figure:setup1}.
While we are only interested in the classical part of this field as it is used solely for the trapping, we include both the quantum and the classical part of the cavity field consisting of a standing wave in $z$-direction and a Gaussian profile in $x$- and $y$-direction,
\be
\begin{split}
\Eop_{\rm cav}(\bx,t)&=\im \sqrt{\frac{\omega_0}{\epsilon_0V_0}}(f_{\rm cav}(\bx)\aop_0-\hc)\\
\mathcal{E}_{\rm cav}(\bx,t)&=\im \sqrt{\frac{\omega_0}{\epsilon_0V_0}}(f_{\rm cav}(\bx)\alpha-\hc)
.\label{eq:ecav}
\end{split}
\ee
This equation denotes the cavity field in the displaced form, where $|\alpha|^2=n_{\rm ph}$ is the mean number of photons in the steady state, $n_{\rm ph}=\sqrt{2P_c\kappa/\omega_0}/(\im \delta+\kappa)$, with $P_c$ being the power of the driving laser. The mode function is given by $f_{\rm cav}(\bx)=\exp\left(-\frac{x^2+y^2}{W_0^2}\right)\cos(k_{0,z}z-\varphi)$, where $\varphi$ denotes the equilibrium position of the dielectric, $\bk_0$ the wave vector of the cavity light, and the mode volume is given by $V_0=L\pi W_0^2/4$ with $L$ being the cavity length and $W_0$ its waist. 
While the classical term merely yields a shift of the trapping frequency and the equilibrium position, the quantum part of the mode function is used to manipulate the cm degree of freedom of the dielectric including the part describing the opto-mechanical coupling. Note that we only consider one mode of the cavity here, higher harmonics supported by the cavity are not included, they are contained in the continuum of homogeneous modes and coupling to them is treated as losses. In case of using a second cavity mode for the trapping instead of the tweezer, Eq.~\eqref{eq:ecav} simply has to be summed over several modes with different profiles. 

The full dynamics of the system is obtained taking into account the coupling of the tweezers and the cavity mode to the vacuum modes, given by Eq.~\eqref{eq:freefield}. The full master equation is described by Eq.~\eqref{eq:mastertot} with the corresponding decay rates given by Eq.~\eqref{eq:decmech} and Eq.~\eqref{eq:kappa}, where $\Gamma$ contains contributions of the cavity mode and the tweezers. 

The specific description of spheres is eased by the availability of an analytical solution, the Mie solution, for the scattered fields and cross sections~\cite{bohrenbook, hulstbook}. 
The Mie solution is based on expanding the incoming electromagnetic field in spherical waves. This expansion suits the sphere's geometry and it is thus possible to apply boundary conditions to determine the scattered fields. Note that while the polarization of the electromagnetic field has been neglected to ease the notation in the previous sections, we take it into account here. The Mie solution is defined for plane-wave states, and we use the relation 
\be
\mathcal{T}_{\bk, \bk'}=\frac{\im c^2}{2\pi\omega_{\bk}}f(\bk, \bk')\label{eq:scattampl}
\ee
to formulate the solution for the cavity field in terms of classical amplitudes.

\begin{figure}
 \includegraphics[width=.49\linewidth]{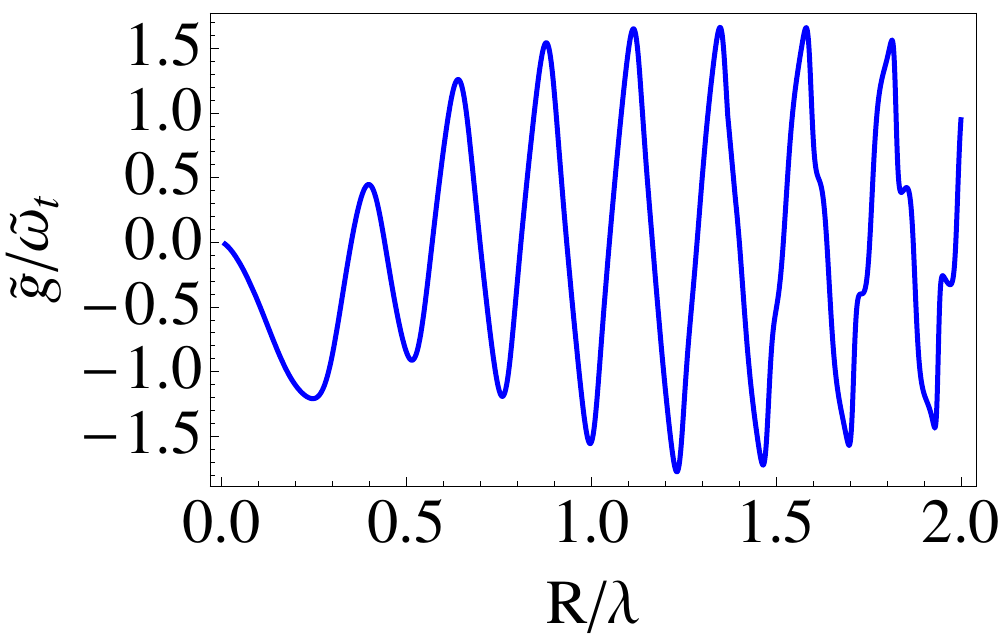}
 \includegraphics[width=.49\linewidth]{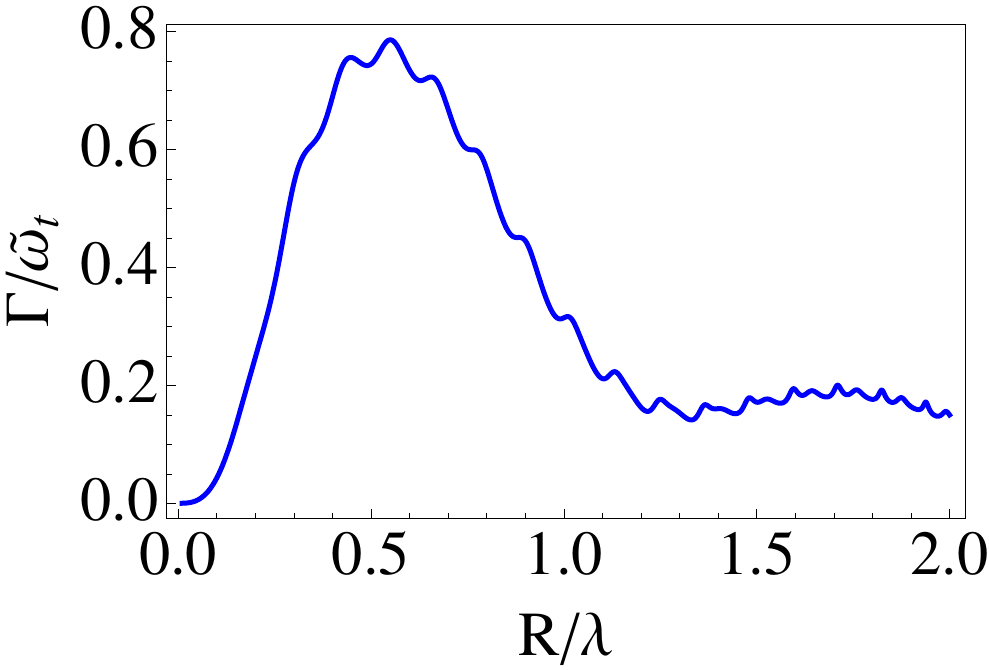}
  \includegraphics[width=.49\linewidth]{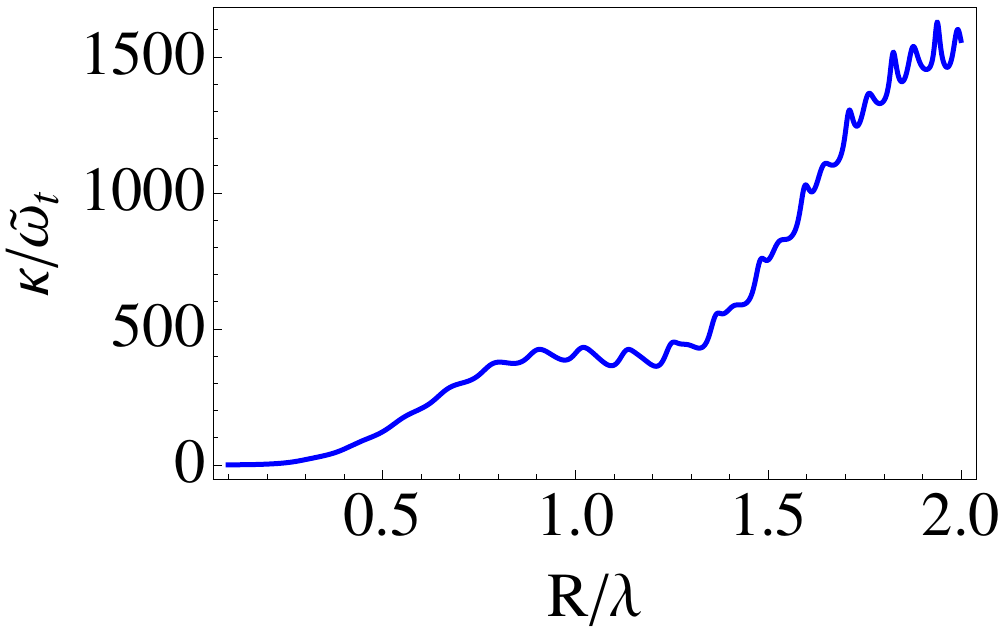}
 \includegraphics[width=.49\linewidth]{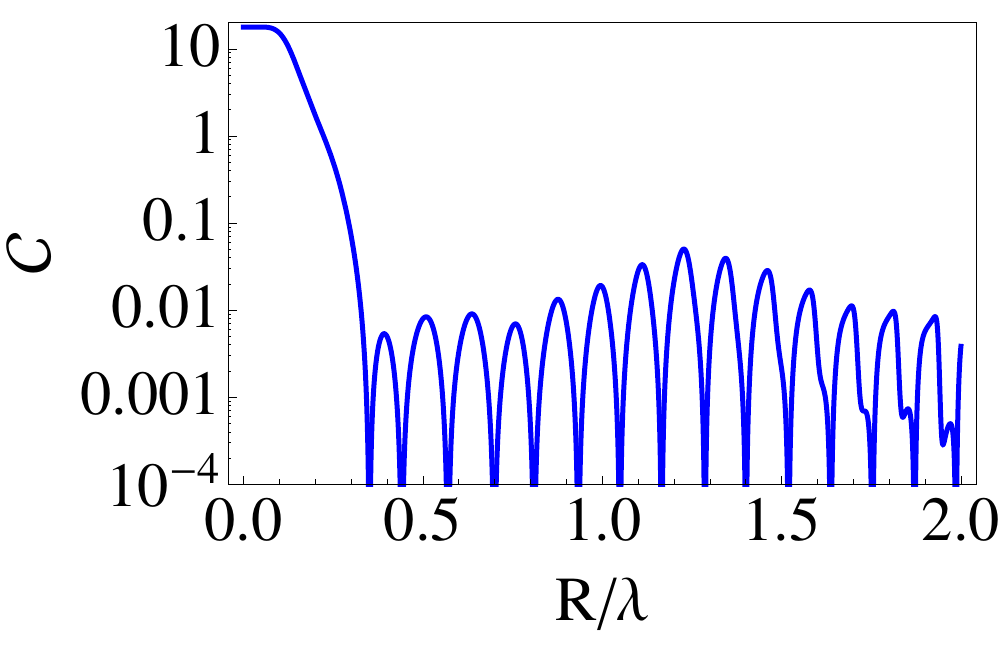}
\caption {Cavity-Optomechanical parameters for different sizes of the object: \textit{Upper left panel:} Optomechnical coupling $\tilde{g}$, \textit{Upper right pannel:} Recoil heating of the center-of-mass $\Gamma$, \textit{Lower left pannel:} Cavity decay rate $\kappa$, \textit{Lower right pannel:} Cooperativity $\mathcal{C}$}
\label{Fig:sphere_R}
\end{figure}

For spherical objects, assuming a vanishing absorption, $\Im[\epsilon_r]\approx 0$, it is possible to connect all quantities to the classical scattering amplitude in the forward-direction using the optical theorem, Eq.~\eqref{eq:opt_th}, which yields 
\be
f(\bk,\bk)=\frac{\sqrt{2\pi}}{|\bk|}\sum_{n=1}^{\infty}(2n+1)(a_n+b_n).
\ee
The coefficients $a_n, b_n$ depend on the dielectric constant and the radius of the sphere, they are defined in terms of spherical Bessel functions. We refer the reader to standard textbooks~\cite{bohrenbook, hulstbook} for their specific form. Hence, the optomechanical parameters can be determined: 
\begin{enumerate}

\item The optomechanical coupling is defined by
\be
\begin{split}
\tilde{g}=&\alpha\frac{x_{0}\pi c}{2 k_0 V_c}\sin(2\varphi)\times\\
&\times\Im\left[\sum_{n=1}^{\infty}(2n+1)(-1)^{n+1}(a_n+b_n)\right],
\end{split}
\ee
where $c$ denotes the velocity of light and $\varphi$ the position of the sphere in the cavity. 
\item The total cavity decay rate is defined by $\kappa_{\rm tot}=\kappa_0+\kappa$, where $\kappa_0$ is the intrinsic cavity decay rate resulting from imperfections in the mirrors and
\be
\begin{split}
\kappa=&\frac{c\pi}{2 k_0^2 V_c}\Re\Large[\sum_{n=1}^{\infty}(2n+1)\times\\
&\times(1+(-1)^{n}\cos(2\varphi))(a_n+b_n)\Large].
\end{split}
\ee
\item The recoil heating of the dielectric due to scattering of cavity photons can be computed as
\be
\begin{split}
 \Gamma^{\rm cav}=& \frac{x_0^2c\alpha^2\pi}{ V_c}\Re[\sum_{n=1}^{\infty}(2n+1)\times\\
 &\times(1+(-1)^n\cos(2\varphi))(a_n+b_n)].
\end{split}
\ee
Note that the recoil heating from the trapping lasers can be obtained in full analogy by inserting the tweezer mode. 
\end{enumerate}
Besides the minimal phonon number $n_{\rm min}$ describing the possibility to cool the system (close) to its quantum-mechanical ground state, another figure of merit to describe the cavity-optomechanical properties is the cooperativity $\mathcal{C}$. This measure for the coherent coupling between the motion and light is defined by
\be
\mathcal{C}=\frac{\tilde{g}^2}{\Gamma\kappa}\label{eq:coop}
\ee
and depends on the size of the particle and its position in the cavity. It is in particular essential to have a sufficiently high cooperativity to perform protocols coupling the cm to the light~\cite{romeroisart2011a}. Assuming that the object is positioned at the maximal slope of the standing wave and is much smaller than the laser's waist, the asymptotic form of the cooperativity is given by
\be
\mathcal{C} \propto \left\{ \begin{array}{ll}
         1/\epsilon_c^2k_0^6R^6 & \mbox{if$ R\ll \lambda$};\\
       1/k_0^2R^2 & \mbox{if $R\gg\lambda$}
        \end{array} \right.\ee
under the assumption that the laser's waist is larger than the object. In case the dielectric is not fully covered by the laser's waist, the beam's Gaussian shape has to be taken into account~\cite{gouesbet09} leading to an even lower cooperativity. In the proceeding, both the minimal phonon number and the cooperativity are used to quantify the system's performance as an optomechanical setup. 
The optomechanical parameters are thus determined as illustrated in Fig.~\ref{Fig:sphere_R} for varying sphere sizes. The experimental parameters are chosen as follows:
\begin{itemize}
\item {\bf Dielectric object:} We assume spheres of fused silica with density $\rho=2201~\rm kg/m^3$, a dielectric constant $\Re[\epsilon_r]=2.1$ and $\Im[\epsilon_r] \sim 2.5 \times 10^{-10}$. We vary their radii between $R=10\rm{nm}-2\mu\rm{m} $ and position them at the maximal slope of the cavity field, $\varphi=\pi/4$.
\item {\bf Cavity:} We assume a confocal high finesse cavity of length $L=4~\rm mm$ and finesse $\mathcal{F}=5 \times 10^5$ leading to a cavity decay rate $\kappa_0=c\pi/2\mathcal{F}L=2\pi \times 44\rm kHz$. This cavity is impinged by a laser of power $P_c=0.1~\rm mW$, wavelength $\lambda=1064~\rm nm$, which gives a waist of $W_0=\sqrt{\lambda d/2\pi}\approx 26~\mu \rm m$

\item {\bf Optomechanical parameters:} The tweezers are constructed with a laser of wavelength $\lambda=1064~\rm nm$ and a lense of high numerical aperture. They supply a harmonic trap for the object of frequency $\tilde{\omega}_{\rm t}=2\pi\times 136~\mathrm{kHz}$ in the transversal direction and a slightly smaller one in the direction of light propagation. The cavity photons have a frequency $\omega_0=2 \pi \times 2.8 \times 10^{14}$ Hz and the steady state photon occupation is $|\alpha|^2\approx 3.7 \times 10^8$.  
\end{itemize}

Considering the graphical illustration of the optomechanical parameters in Fig.~\ref{Fig:sphere_R}, the absolute value of the optomechanical coupling $\tilde{g}$ first increases with the radius $R$ reaching a local maximum at $R\approx 260~\rm nm$, then decreases and even vanishes at $R\approx 370~\rm nm$. In the following it continues these oscillations. The decoherence rate of the cm motion first increases $\propto R^3$, then begins to fall off for $R\gtrsim600\rm~nm$. This is due to its dependence on the ground-state size and the cross section, where the scattering is described by the Rayleigh cross section $\propto R^6$ for small objects, to give way to a scaling $\propto R^2$ in the limit of geometrical scattering and the squared ground state size, which is $\propto R^{-3}$. Also the cavity decay rate increases $\propto R^6$ at first, then shows some plateaus to finally converge to a scaling $\propto R^2$. Consequently, the cooperativity first decreased immensely to exhibit oscillations later on. These oscillations can only be predicted taking multiple-scattering processes into account. Nevertheless, the maximal values of the cooperativity are merely $\mathcal{C}\approx 0.05$. The minimal phonon number is obtained by minimizing the function $\bar{n}$ described in Eq.~\eqref{eq:nsteady} with respect to $\delta$. While ground-state cooling is feasible for spheres $R\lesssim 250~\rm{nm}$, only relatively large final phonon numbers can be achieved for larger spheres, \eg, $n_{\rm min}\approx 350$ for $R\approx1.3\mu \rm{m}$.

 \begin{figure}
 \includegraphics[width=.9\linewidth]{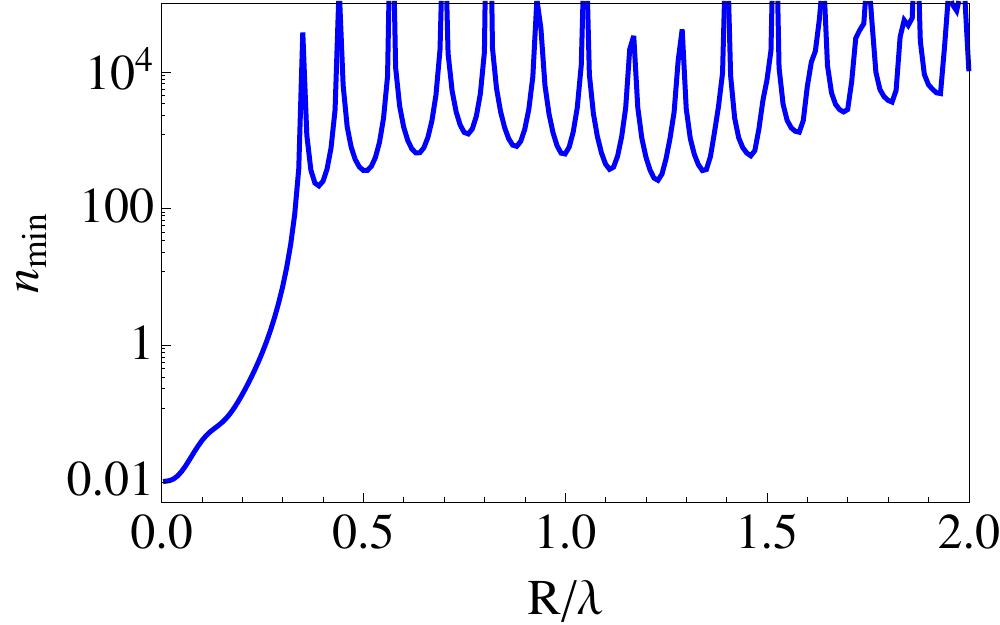}
\caption{Minimal phonon number attainable for different sphere sizes with the experimental parameters given in the main text.}
\label{Fig:nmin_phon}
\end{figure}

\section{Conclusion and Outlook}\label{sec:conclusion}

We have derived a full quantum theory to describe the coupling of light to the center-of-mass motion of non-absorbing dielectrics of arbitrary sizes and shapes in optical cavities. 
The common approach to the description of sub-wavelength dielectrics and light is within the framework of Born-Markov master equations~\cite{gardinerbook}. For quantum-optical systems, interactions among different bath modes are commonly not present or negligible. However, as the size of the dielectric is increased to the order of the cavity wavelength, this approximation is no longer justified. In this article, an approach to take these interactions into account has been developed. This is achieved by describing the bath operators of the electromagnetic field in terms of their scattering solution in the presence of a dielectric and developing a master equation in this new basis. Based on this method, it is possible either to solve the problem exactly, if analytical solutions are available, or to truncate the series of multi-scattering processes in a controlled way. 
The resulting master equation of Lindblad form gives a full description of the system's dynamics, more specifically, decoherence rates for both the cavity and the light mode as well as renormalizations to the Hamiltonian can be calculated. 

In the second part of the paper, we apply these methods to optically levitating spheres of arbitrary size, determining all optomechanical parameters. In particular, the solutions are expanded in terms of classical scattering amplitudes for plane-wave states, demonstrating that the knowledge of these amplitudes is sufficient to fully analyze the system. In the spherical case, an analytical solution for the scattering amplitudes, the Mie solution, simplifies computations. Summarizing the results, we find that on the one side, small spheres can be cooled to their quantum-mechanical ground state and can be efficiently addressed by light due to a high cooperativity. On the other side, as expected, ground-state cooling is out of reach for larger spheres and only relatively high phonon occupations $n_{\rm min}\approx 350$ can be obtained. 

Albeit the focus of this article is on the general derivation of the theory and the analysis has been restricted to transparent spheres without internal structure, we would like to give an outlook on two extensions of this work that might significantly decrease the minimal phonon numbers.
First, objects which are better-suited, or even tailored to the cavity mode, might perform more efficiently as an optomechanical device. Given the fact that they scatter most photons back into the cavity mode, losses are reduced yielding more benign conditions for optomechanics and thus lower phonon numbers and a higher cooperativity.
Another possibility could be to develop new cooling schemes based on the additional use of internal degrees of freedom of the sphere. As the sphere grows larger, the coupling between the cm and the relative motion is increased and additionally displays some resonances. These resonances hold the potential to be used to cool the cm motion via coupling to the relative degrees of freedom, already being sparsely occupied due to their high frequencies.

\begin{acknowledgments}
The authors thank the IQI at the California Institute of Technology, where part of this work has been carried out, for hospitality. We also thank N.~Kiesel and A.~ Manjavacas for valuable discussions, and J.~P.~Ronzheimer for support with graphical illustrations. Funding by the Elite Network of Bavaria (ENB) project QCCC (A.C.P.), the EU project AQUTE and Caixa Manresa is gratefully acknowledged.
\end{acknowledgments}

\appendix

\section{ The classical approach}\label{classical}
This Appendix sketches the solution of the equations of motion of the classical electrodynamic field, giving an overview of the possibilities and limitations of the description in classical scattering theory. The approach in the classical case is to solve Maxwell's equations~\cite{jacksonbook}. Neglecting polarizations yields
\be
\begin{split}
E(\bold{x},t)&=E_{\rm in}(\bold{x},t)+\epsilon_c \int d\bx' \mathcal{G}(\bold{x}', \bold{x})E(\bold{x}',t),\label{elfield_cl}
\end{split}
\ee
where $E_{\rm in}(\bold{x},t)$ denotes the incoming electromagnetic field and $\mathcal{G}(\bold{x}', \bold{x})=|\bk_0|^2\exp(\im |\bk_0| |\bx-\bx'|)/|\bx-\bx'|$ the propagator ($\bk_0$ being the wave vector of the incoming field). 
This self-consistent equation has been intensely studied in classical scattering theory and is in general only solvable approximately. 
There exist only few geometries, like, \eg , a cylinder, a sphere, or an ellipsoid, where analytical solutions are tractable. In the special case of a spherical object, the scattered electric field can be determined exactly by expanding the field in spherical waves and subsequently applying boundary conditions, yielding the so-called Mie solution~\cite{strattonbook, bohrenbook, hulstbook}. Perturbative approaches~\cite{Yeh64, Erma68}, based on the analytical solution and an extension of the treatment via distorting the surfaces at different points, only allow for calculations of small perturbations. Numerical approaches like the discrete dipole ansatz~\cite{purcell73} or the T-matrix method (see, \eg,~\cite{Stroem75} for an expository article) are applicable to a larger class of objects and are widely used today.
Indeed, these approaches coincide with the analytical solution for perfectly spherical objects~\cite{purcell73}. In the limit of very large objects, $R\gg \lambda$, applying a ray-optics approach immensely simplifies the calculation of forces on the dielectric~\cite{ashkin92}. A further analysis of the classical solution is beyond the scope of this article and we refer the reader to the literature, for example,~\cite{abajo07} for a more detailed discussion.

Once the electromagnetic field including the scattering is obtained, the classical radiation force is determined via the momentum conservation law: the force acting on the dielectric is the change in momentum of the EM field and can be determined from Maxwell's stress tensor. 
The total force on an object interacting with the EM field consists of the change of mechanical momentum and field momentum $\bold{F}_{\rm tot}=d\bold{P}_{\rm mech}/dt+d\bold{P}_{\rm field}/dt$ with
\be
\begin{split}
\frac{d}{dt}\bold{P}_{\rm mech}&=\int_Vd\bx(\rho_{e} \bold{E}+\bold{J}\times \bold{B})\\
\frac{d}{dt}\bold{P}_{\rm field}&=\frac{d}{dt}\int_Vd\bx(\bold{E}\times\bold{B}),
\end{split}
\ee
where $\bold{B}$ denotes the magnetic field, $\rho_{e}$ the charge density and $\bold{J}$ the current.
Rewriting and manipulating this equation (see~\cite{jacksonbook} for details) yields a formulation in terms of Maxwell's stress tensor,
\be
\bold{F}_{\rm tot}=\int_S \bold{T}\bold{n}dA,
\ee
where the integration is taken over the surface $dA$ of the object and $\bold{n}$ is the outward normal vector to the closed surface $S$. Maxwell's stress tensor is given by the electric and magnetic fields
\be
T_{\alpha \beta}=\epsilon_0\left[E_{\alpha}E_{\beta}+c^2B_{\alpha}B_{\beta}-\frac{1}{2}(\bold{E}\bold{E}+c^2\bold{B}\bold{B})\delta_{\alpha \beta}\right].
\ee
Plugging in the expression for the scattered electromagnetic field in the above equation, we can determine the forces on the dielectric.
This method to determine forces enables one to calculate trapping of dielectrics and also to calculate radiation pressure effects. However, this approach cannot be used to determine a full dynamical description of the system and its decoherence rates.

\section{Wigner-Weisskopf with correlations in the field}\label{sec:wigner}
In this Appendix, an alternative approach to the description of the interaction between a single photon and a dielectric is given within the Wigner-Weisskopf ansatz. In contrast to a direct description with master equations, where information about the light fields can be extracted via the quantum regression theorem, the Wigner-Weisskopf approach directly yields expressions for the photonic fields. Analyzing the light emitted from the cavity yields information about the mechanical state of the system. Following~\cite{cirac93}, it is possible to determine the occupation of the mechanical mode, and thus, to monitor the cooling of the system. Complementing the master equation ansatz, in App.~\ref{Schrodinger}, we solve the equations of motion for the coefficients of the density matrix taking into account correlations among the free modes of the field. In full analogy to the solution of the Heisenberg equations of motion given in Sec.~\ref{sec:freefield}, the equations of motion for the coefficients of the density matrix can be demonstrated to be equivalent to the Lippmann-Schwinger equation enabling one to use solutions of the classical scattering equations. Subsequently, an inhomogeneity in the electromagnetic field is added in App.~\ref{der:cavity} and its effect on the scattered fields is investigated. Finally, a master equation describing the decay of the cavity mode is derived in App.~\ref{coeff_master}.

\subsection{Free photons}\label{coefficient}\label{Schrodinger}
Here, the evolution of a single photon in a plane-wave state interacting with a dielectric is discussed. For simplicity, the motion of the object is neglected for now, assuming $M\rightarrow \infty$. 
The coefficients of the wave function in the Schr\"odinger picture are defined by
\be
\begin{split}
|\psi\rangle&=\int d\bk c_{\bk}(t)\adop_{\bk}|\Omega\rangle\label{eq:state},
\end{split}
\ee 
where $|\Omega\rangle$ denotes the vacuum state. The assumption that the object's mass is infinite manifests itself in the independence of the wave function of the object's momentum state: the effect of the photon's recoil on the dielectric is neglected. 
To obtain the Eqs. of motion, we let the homogeneous part of the Hamilton given by Eq.~\eqref{eq:bath}, $\hat{H}_{\rm B}$, act on the above wave function. This yields%
\be
\begin{split}
\im \dot{c}_{\bk}(t)=&\omega_{\bk} c_{\bk}(t)
+ \epsilon_c\int_{V(\br)}d\bx\int d\bk'\sqrt{\omega_{\bk}\omega_{\bk'}}\\
&\times(c_{\bk'}(t')e^{\im(\bk-\bk')\bx}+c^*_{\bk'}(t)e^{\im(\bk+\bk')\bx}).\label{eq:coeff}
\end{split}
\ee
In order to close this system of equations, we define
\be 
h^{(+)}(\bold{x}, \br, t)=\im\int d\bk \sqrt{\omega_{\bk}}e^{-\im \bk (\bx+\br)}c_{\bk}(t),\label{fourierb}
\ee
with $h(\bold{x}, \br, t)=h^{(+)}(\bold{x}, \br, t)+h^{(-)}(\bold{x}, \br, t)$. Subsequently we multiply both sides of Eq.~\eqref{eq:coeff} by $\im \int d\bk \sqrt{\omega_{\bk}}\exp(-\im \bk \bx)$.
A formal integration over time and a transition to the frame rotating at a frequency $\omega_0$, $\tilde{h}(\bold{x},\br, t)^{(\pm)}=\exp(\pm\im\omega_{0}t)h^{(+)}(\bold{x},t)$, yields
\be
\begin{split}
h^{(+)}(\bold{x}, \br, t)=&h^{(+)}_{\rm in}(\bold{x}, \br, t)+\epsilon_c \int d\bk\int_{V(\br)} d\bx'e^{-\im \bk (\bx-\bx')}\\
&\times \frac{\omega_{\bk}}{\omega_{\bk}-\omega_0+\im \gamma}h^{(+)}(\bold{x}',\br, t),\label{eq:motionb}
\end{split}
\ee
where $h^{(+)}_{\rm in}(\bold{x}, \br, t)$ is defined in analogy to Eq.~\eqref{eq:ein}. Also here, we assume $h^{(+)}(\bold{x}, \br, t)$ to be peaked at $\omega_0$.
In order to solve this differential equation, we have taken the slowly-varying approximation, assuming that $\tilde{h}(\bold{x}, \br, t)^{(\pm)}$ can be taken out of the integration that is extended to $t\rightarrow \infty$.
A transformation back to the coefficient picture thus gives in full analogy to the operators of the electromagnetic field, Sec.~\ref{sec:freefield},
\be
\begin{split}
c_{\bk}(t)=&\int d\bk'e^{-\im\omega_{\bk'}t}\times\\
&\times\left(\delta(\bk-\bk')+\frac{\mathcal{T}_{\bk, \bk'}(\br)}{\omega_\bk'-\omega_0+\im \gamma}\right)c_{\bk'}(0).\label{eq:cvak}
\end{split}
\ee
The coefficients contain the information about the full dynamical evolution of the system and can be used to reconstruct its density matrix.

\subsection{Cavity field}\label{der:cavity}
In this section the analysis of the previous chapter is extended to the more general case, where an inhomogeneity in the electromagnetic field is present. This inhomogeneity is typically a cavity that changes the system's mode distribution. Also in this case, the cm degree of freedom is treated as a number and its motion in the cavity is neglected. We solve the Schr\"odinger eqs. of motion for the coefficients in Sec.~\ref{solv_inhom} and subsequently derive the master equation for the time evolution of the cavity mode $\aop_0$ in Sec.~\ref{coeff_master}. 
\subsubsection{Solution of the inhomogeneous part}\label{solv_inhom}
The wave function including the cavity mode is written as
\be
|\psi\rangle=c_0(t)\adop_0|\Omega\rangle+\int d\bold{k}c_{\bk}(t)\adop_{\bk}|\Omega\rangle+s_0|\Omega\rangle,\label{eq:wave}
\ee
where $s_0$ is constant and $c_0(t)$, $c_{\bk}(t)$ are time-dependent. 
The Hamiltonian, Eq.~\eqref{eq:hamdi} causing the scattering consists of two parts, $\hat{W}(\br)$, Eq.~\eqref{eq:bathint} describing the coupling among the homogeneous modes and $\hat{H}_{\rm BS}$, Eq.~\eqref{eq:hint} denoting the coupling of the inhomogeneity (cavity mode) to the free ones. The following equations of motion are obtained:

\be
\begin{split}
\im \dot{c}_{\bold{k}}(t)=&\epsilon_c\int_{V(\br)} d\bold{x}\int d \bold{k}'e^{\im(\bold{k}-\bold{k}')\bold{x}}\sqrt{\omega_{\bk}\omega_{\bk'}}c_{\bold{k}'}(t)\\
&+\epsilon_c\int_{V(\br)} d\bold{x}\sqrt{\frac{\omega_{\bk}\omega_0}{2 V_0}}e^{-\im \bold{k}\bold{x}}f(\bx)c_0(t)+\omega_{\bk}c_{\bold{k}}(t)\label{eq:call}
\end{split}
\ee
and
\be
\begin{split}
\im \dot{c}_0(t)=&\epsilon_c\int_{V(\br)}d\bold{x}\int d\bold{k}\sqrt{\frac{\omega_0\omega_{\bk}}{2V_0}}e^{\im \bold{k\bold{x}}}f^*(\bold{x})c_{\bold{k}}(t)\\
&+\omega_0c_0(t)\label{eq:czero},
\end{split}
\ee
where the rotating-wave approximation has been taken, which is equivalent to remaining in the one-excitation manifold.
Using Eq.~\eqref{fourierb}, we can simplify Eqs.~\eqref{eq:call},~\eqref{eq:czero} to obtain after an integration over time
\be\label{eq:binhom}
h^{(+)}_{\rm inh}(\bold{x},\br, t)=h^{(+)}(\bold{x},\br, t)+d(\bx,\br, t)
\ee
with
\be
\begin{split}
d(\bold{x},\br, t)&=\int d\bold{k}\mathcal{A}(\bold{k},\br, t)e^{\im \bold{k}\bold{x}},
\end{split}
\ee
and
\be
\begin{split}
\mathcal{A}(\bk,\br, t)=& \epsilon_c\int_0^td\tau e^{-\im(\omega_0-\omega_{\bold{k}})\tau}\times\\
&\times\int_{V(\br)} d\bold{x}'\omega_k\sqrt{\frac{\omega_0}{2 V_0}}e^{-\im \bold{k}\bold{x}'}f(\bold{x}')c_0(\tau).
\end{split}
\ee
The first part of the integration in Eq.~\eqref{eq:binhom} has been carried out under the Markov assumption, which is justified as correlations in the electromagnetic field decay quickly. No approximation is taken for the time evolution of the inhomogeneous part and it is kept in the most general form for now. The strategy to find a solution for the inhomogeneous case described by Eq.~\eqref{eq:binhom} is to connect it to the homogeneous one, described in the previous section, Sec.~\ref{Schrodinger}.

The solution of the homogeneous case, Eq.~\eqref{eq:motionb}, can be formally written in vector-form as
\be
\begin{split}
\bold{h}&=\bold{h}_{\rm{in}}+\hat{\mathcal{B}}\bold{h}\\
\bold{h}&=\frac{1}{1-\hat{\mathcal{B}}}\bold{h}_{\rm{in}}.\label{eq:vecs},
\end{split}
\ee
where $\hat{\mathcal{B}}$ describes the scattering operator in matrix form and $\bold{h}$ denotes the continuous vector-representation of $h(\bx,\br, t)$.
Comparing the homogeneous case to the inhomogeneous one, an additional inhomogeneous term is present 
leading in analogy to Eq.~\eqref{eq:vecs} to
\be
\begin{split}
\bold{h}&=\bold{h}_{\rm{in}}+\hat{\mathcal{B}}\bold{h}+\bold{d}\\
\bold{h}&=\frac{1}{1-\hat{\mathcal{B}}}(\bold{h}_{\rm{in}}+\bold{d})\label{eq:vecsinhom},
\end{split}
\ee
where $1/(1-\hat{\mathcal{B}})$ denotes the solution-operator for the plane-wave state. This equivalence facilitates the solution of the inhomogeneous system by connecting it to the homogeneous one. 
The system is initially assumed to have one photon in the cavity mode and none in the homogeneous modes, $c_{\bk}(0)=0$, such that $\bold{h}_{\rm in}=0$ and Eq.~\eqref{eq:binhom} can be solved as
\be
\begin{split}
h^{(+)}(\bold{x},\br,t)=&d_{\rm in}(\bx,\br,t)+\epsilon_c\int d\bold{k}\int_{V(\br)}d\bx'e^{-\im \bk(\bx-\bx')}\\
&\times \frac{\omega_{\bk}}{\omega_{\bk}-\omega_0+\im\gamma}d(\bx,\br, t),\label{eq:binh}
\end{split}
\ee
where $d_{\rm in}(\bx,\br,t)$ is defined in analogy to Eq.~\eqref{eq:ein}.
Taking the inverse transformation yields the solution for the coefficients $c_{\bk}(t)$
\be
\begin{split}
c_{\bk}(t)=&\int d\bk'e^{-\im\omega_{\bk'}t}\times\\
&\times\left(\delta(\bk-\bk')+\frac{\mathcal{T}_{\bk, \bk'}(\br)}{\omega_{\bk'}-\omega_0+\im \gamma}\right)\mathcal{A}(\bk',\br,0).\label{eq:cvakinh}
\end{split}
\ee
Plugging Eq.~\eqref{eq:binh} back into Eq.~\eqref{eq:czero}, we can close the equations of motion. After taking the Markov approximation assuming that the system does not change significantly during the interaction with the environment, such that $c_{ 0}(t-\tau)\approx c_{0}(t)$. Using some standard relations for the scattering operators, the time evolution of the inhomogeneous mode in terms of transition operators is given by
\be
\frac{\dot{c}_0(t)}{c_0(t)}=-\int d\bk |\mathcal{T}_{\bk, \rm c}(\br)|^2\left[\pi\delta(\omega_{\bk}-\omega_0)+\im \mathcal{P}\frac{1}{\omega_{\bk}-\omega_0}\right].\label{eq:cdot_trans}
\ee
The effect on the light field can be determined approximating $\mathcal{T}_{\bk, \rm c}(\br)\approx\mathcal{T}_{\bk, \rm c}(0)$ thus neglecting the effect on the cm mode.
Integration gives
\be
c_0(t)=\exp((-\kappa+\im\Delta^{\rm L})t)c_0(0)\label{eq:c0}
\ee
with the decay rate
\be
\begin{split}
&\kappa=-\Re\left(\frac{\dot{c}_0(t)}{c_0(t)}\right)\label{eq:kappa1}
\end{split}
\ee
and a Lamb shift
\be
\begin{split}
&\Delta^{\rm L}=-\Im\left(\frac{\dot{c}_0(t)}{c_0(t)}\right).
\label{eq:delta1}
\end{split}
\ee
These results are in accordance with Fermi's Platinum Rule, the extension of Fermi's Golden Rule to all orders in multiple-scattering processes.

\subsubsection{From the coefficients to the master equation}\label{coeff_master}
Starting from the wave function Eq.~\eqref{eq:wave}, the system's density matrix is obtained by tracing out the environment and can be written as~\cite{breuerbook}
\be
\hat{\rho}_S(t)=\begin{pmatrix} |c_0(t)|^2&s_0^*c_0(t)\\ s_0c_0(t)^*&1-|c_0(t)|^2
\end{pmatrix}.
\ee
Taking the derivative gives
\be
\dot{\rho}_S(t)=\begin{pmatrix} \frac{d}{dt}|c_0(t)|^2&s_0^*\dot{c}_0(t)\\ s_0\dot{c}_0(t)^*&-\frac{d}{dt}|c_0(t)|^2
\end{pmatrix}.
\ee
Using Eqs.~\eqref{eq:kappa1}, \eqref{eq:delta1} finally yields
\be
\dot{\rho}_{\rm S}=-\im\left[\Delta^{\rm L} \aop_0^{\dagger}\aop_0, \hat{\rho}_{\rm S}\right]
+\mathcal{L}^{\rm L}[\hat{\rho}_{\rm S}]\label{eq:inhom}
\ee
with 
\be
\mathcal{L}^{\rm L}[\hat{\rho}_{\rm S}]=\kappa\left(2\aop_0\hat{\rho}_{\rm S}\aop_0^{\dagger}-[\aop_0^{\dagger}\aop_0,\hat{\rho}_{\rm S})]\right),\label{eq:Lcav}
\ee
where $\Delta^{\rm L}$ denotes a shift of the energy levels, and $\kappa$ describes the decay rate of the cavity photons due to losses effected by the presence of the dielectric. This master equation and its decay rates are equivalent to the result for the light fields obtained in Sec.~\ref{sec:master}.

\section{Contributions to the master equation in the Lamb-Dicke regime}\label{master_opto}
In this Appendix, we demonstrate in some detail which terms of the master equations discussed in Sec.~\ref{sec:optomech} can be neglected within the Lamb-Dicke regime under strong driving. We make in particular use of the analytic property of the transition matrix elements and their derivatives. As an example, we discuss the contribution $\propto \br$ given by
\be
\begin{split}
\hat{H}_{\rm rn}^{\rm shift}=&\alpha^2\int d\bk (\delta(\omega_{\bk}-\omega_0)\Im\left[\mathcal{T}'_{ \bk, \rm c}(\br)|_{\br=0}\mathcal{T}^{*}_{ \rm c, \bk}(0)\right]\\
&+\mathcal{P}\frac{1}{\omega_{\bk}-\omega_0}\Re\left[\mathcal{T}'_{ \bk, \rm c}(\br)|_{\br=0}\mathcal{T}^{*}_{ \rm c, \bk}(0)\right])\br.
\end{split}
\ee
Under the assumption that $t(\omega_{\bk})=\mathcal{T}'_{\bk, \rm c}(\br)|_{\br=0}\mathcal{T}^{*}_{\rm c, \bk}(0)$ is analytic in $\omega_\bk$, the Hilbert transformation can be used to show that
\be
\begin{split}
\int d\omega_{\bk}\mathcal{P}\frac{1}{\omega_{\bk}-\omega_0}\Re\left[t(\omega_{\bk})\right]=-\Im\left[t(\omega_{0})\right]
\end{split}
\ee
and consequently $\hat{H}_{\rm rn}^{\rm shift}=0$. This transformation is also used to simplify the renormalization contribution $\propto \br^2$ in Sec.~\ref{sec:optomech}. 

Another argument holds when neglecting the incoherent contribution arising from terms $\propto \aop_0\alpha^* \br$. This part describes decoherence of the mechanical and the light degree of freedom
\be
\begin{split}
\mathcal{L}^{\rm g}[\rho_{\rm S}]=&2\Gamma_g(\aop_0\rhos(\bop+\bdop)-[\aop_0(\bop+\bdop),\rhos])\\
&+2\Gamma_g^*((\bop+\bdop)\rhos\adop_0-[\adop_0(\bop+\bdop),\rhos]),\label{eq:decg}
\end{split}
\ee
where 
\be
\Gamma_g=x_0\alpha^*\int d\bk \delta(\omega_{\bk}-\omega_0)\mathcal{T}^*_{ \rm c, \bk}(0)\mathcal{T}'_{\bk, \rm c}(\br)|_{\br=0}.
\ee
 This contribution can in general be neglected for the cm degree of freedom, as it is suppressed by $1/\alpha$ compared to Eq.~\eqref{eq:deccms}. Requiring that $\alpha \eta \ll1$, we can also neglect the effect of this contribution on the cavity mode. This requirement becomes clear comparing Eq.~\eqref{eq:decg} to the decay of the cavity described by Eq.~\eqref{eq:deccav}.

\section{The small particle limit}\label{limits}
In this section, we consider the important limit of the general theory, where the dielectric is smaller than the cavity wavelength. In this case, it is justified to neglect couplings among different bath modes as described by Eq.~\eqref{eq:bathint}. While this derivation has been analyzed in~\cite{romeroisart2011a, nimmrichter10}, we demonstrate in this Appendix that the same master equation can be obtained starting from the general description, Eq.~\eqref{eq:master}, and considering only terms of the lowest order in the Born series of scattering theory. Let us now discuss the different contributions and how they are modified:
\begin{enumerate}
\item All terms in $\hat{H}_S$ remain the same, the trapping frequency $\tilde{\omega}_{\rm t}$ can be determined from Eq.~\eqref{eq:Hsys}, which for objects much smaller than the wavelength is given by
\be
\omega_t^{0}=\sqrt{\frac{4  \epsilon_{\rm c}}{\rho c} \frac{I}{W_t^2}},
\ee
where $\rho$ is the material's density, $I$ the laser intensity of the optical tweezer and $W_{\rm t}$ its waist.
Also the optomechanical coupling $g$ is given by Eq.~\eqref{eq:Hsys} simplifying to
\be
g^{0}=-x_{0} \frac{\epsilon_{\rm c}\omega_c^2 V }{ 4 c V_c},
\ee
where $V$ is the object's volume. Note that the expressions for $\hat{H}_{\rm S}$ are not affected by neglecting the coupling among the bath modes as these quantities are determined only by system operators.
\item The renormalization of the optomechanical coupling and the trapping frequency are obtained by considering only the lowest order of the Born series $\mathcal{T}_{\bk, \rm c}(\br)\approx\mathcal{V}_{\bk, \rm c}(\br)$ in the expressions for $\Delta$ and $g_{\rm rn}$ given by Eqs.~\eqref{eq:delta},~\eqref{eq:gopto}.
This gives
\be
g_{\rm rn}^{0}=-\epsilon_ck_0^2R^2g^{0}
\ee
with $R$ being the sphere's radius. The renormalization of the trapping frequency is obtained by inserting the trapping mode and leads to 
\be
\Delta^{\rm M,0}=-\epsilon_ck_0^2R^2\omega_t^{0}.
\ee
\item The same procedure, namely taking the lowest order in the Born series by setting  $\mathcal{T}_{\rm c, \bk}(\br)\approx\mathcal{V}_{\rm c, \bk}(\br)$ is also applied to obtain the decoherence rates. For the cavity decay rate this gives
\be
\begin{split} \label{eq:kscs}
\kappa^{0}&=\frac{\epsilon_c^2k_0^4V^2c}{24 \pi  V_c}
\end{split}
\ee
and for the recoil heating of the cm we obtain
\be
\Gamma^{0} =\frac{\epsilon_c^2 k_0^6 V }{6 \pi \rho \omega_t} \left( \frac{P_t}{\omega_L \pi W_t^2} + \frac{ n_\text{ph}c }{2 V_c} \right),
\ee
\end{enumerate}
where the first term describes decoherence due to recoil heating by photons from the tweezers and the second term by photons of the cavity mode. 
Comparison to~\cite{romeroisart2011a} shows that this result is in agreement with the what is obtained when directly taking the Born Markov approach, neglecting interactions between bath modes and deriving the master equation in the standard way. (Note that there are some differences in prefactors as we take into account the polarization of the light here.)


\end{document}